\documentclass{aastex631}
\usepackage{psfig,epsfig,amsfonts,amsmath,amssymb,graphicx,morefloats,appendix,tensor}
\usepackage{color}
\usepackage{natbib}
\usepackage{hyperref}
\usepackage{subfigure}
\renewcommand{\dataset}[1]{doi:{#1}, \url{https://doi.org/#1}}

\begin{document}

\title{Preparing for the Early eVolution Explorer: Characterizing the photochemical inputs and transit detection efficiencies of young planets using multiwavelength flare observations by TESS and Swift}

\author[0000-0002-0583-0949]{Ward S. Howard}
\affil{Department of Astrophysical and Planetary Sciences, University of Colorado, 2000 Colorado Avenue, Boulder, CO 80309, USA}
\affil{NASA Hubble Fellowship Program Sagan Fellow}

\author[0000-0001-7891-8143]{Meredith A. MacGregor}
\affil{Department of Physics and Astronomy, Johns Hopkins University, 3400 N Charles St, Baltimore, MD 21218, USA}

\author[0000-0002-9464-8101]{Adina D. Feinstein}
\affil{Laboratory for Atmospheric and Space Physics, University of Colorado Boulder, UCB 600, Boulder, CO 80309}
\affil{Department of Physics and Astronomy, Michigan State University, East Lansing, MI 48824, USA}
\affil{NASA Hubble Fellowship Program Sagan Fellow}

\author[0000-0002-5928-2685]{Laura D. Vega}
\affil{Exoplanets and Stellar Astrophysics Laboratory, NASA Goddard Space Flight Center, Greenbelt, MD 20771, USA}
\affil{Department of Astronomy, University of Maryland, College Park, MD 20742, USA}
\affil{Center for Research and Exploration in Space Science \& Technology (CRESST II), NASA/GSFC, Greenbelt, MD 20771, USA}

\author[0000-0002-3656-6706]{Ann Marie Cody}
\affil{SETI Institute, 339 N Bernardo Ave Suite 200, Mountain View, CA 94043, USA}

\author[0000-0001-8292-1943]{Neal J. Turner}
\affil{Jet Propulsion Laboratory, California Institute of Technology, 4800 Oak Grove Drive, Pasadena, CA 91109, USA}

\author[0000-0002-0267-9833]{Valerie J. Scott}
\affil{Jet Propulsion Laboratory, California Institute of Technology, 4800 Oak Grove Drive, Pasadena, CA 91109, USA}

\author[0000-0002-0040-6815]{Jennifer A. Burt}
\affil{Jet Propulsion Laboratory, California Institute of Technology, 4800 Oak Grove Drive, Pasadena, CA 91109, USA}

\author[0000-0002-4115-0318]{Laura Venuti}
\affil{SETI Institute, 339 N Bernardo Ave Suite 200, Mountain View, CA 94043, USA}
\affil{Visiting Fellow, School of Physics, UNSW Science, Kensington, NSW 2052, Australia}

\begin{abstract}
Ultraviolet flare emission can drive photochemistry in exoplanet atmospheres and even serve as the primary source of uncertainty in atmospheric retrievals. Additionally, flare energy budgets are not well-understood due to a paucity of simultaneous observations. We present new near-UV (NUV) and optical observations of flares from three M dwarfs obtained at 20 s cadence with \textit{Swift} and TESS, along with a re-analysis of flares from two M dwarfs in order to explore the energy budget and timing of flares at NUV--optical wavelengths. We find a 9000 K blackbody underestimates the NUV flux by $\geq$2$\times$ for 54$\pm$14\% of flares and 14.8$\times$ for one flare. We report time lags between the bands of 0.5--6.6 min and develop a method to predict the qualitative flare shape and time lag to 36$\pm$30\% accuracy. The scatter present in optical-NUV relations is reduced by a factor of 2.0$\pm$0.6 when comparing the total NUV energy with the TESS energy during the FWHM duration due to the exclusion of the $T_\mathrm{eff}\approx$5000 K tail. We show the NUV light curve can be used to remove flares from the optical light curve and consistently detect planets with 20\% smaller transits than is possible without flare detrending. Finally, we demonstrate a 10$\times$ increase in the literature number of multi-wavelength flares with the Early eVolution Explorer (EVE), an astrophysics Small Explorer concept to observe young clusters with simultaneous NUV and optical bands in order to detect young planets, assess their photochemical radiation environments, and observe accretion.
\end{abstract}


\section{Introduction}\label{sec:intro}
The evolution of stellar activity from young stars and the formation of planetary systems are inextricably linked. Heightened levels of stellar emission at X-ray to ultraviolet (UV) wavelengths during the pre-main sequence drives atmospheric escape \citep{Owen_Wu:2013, Lopez_Fortney:2014, Ribas:2016} and disequilibrium chemistry in planetary atmospheres \citep{Loyd:2018b, Tilley:2019, Chen:2021}. Atmospheric escape models find enhanced X-ray and extreme UV (XUV) emission during the first 100 Myr of the stellar lifetime is responsible for the majority of atmospheric loss for small planets with radii near 2$R_\oplus$ (e.g. \citealt{Owen_Wu:2013, Jin:2014, Lopez_Fortney:2014}). At the same time, accretion of material from the inner edge of the protoplanetary disk onto the star is thought to drive stellar angular momentum evolution through both spin up (e.g. \citealt{Armitage:1996, Amard:2023, Zwintz:2022}) and disk locking (e.g. \citealt{Koenigl:1991, Venuti:2017, Rebull:2018}). Angular momentum evolution is responsible for changes in the strength and organization of the surface magnetic field with age, leading to a general decrease in stellar activity in the form of spots and flares (e.g. \citealt{West:2008, West:2015, Davenport:2019, Magaudda:2020, Ilin:2021}).

The enhanced rate of stellar flares at young ages provides a significant contribution to the high-energy radiation environments of planets orbiting young stars \citep{Loyd:2018b, Johnstone:2021}. Although true of all pre-main sequence stars, the impact of this activity is thought to be most significant for planets orbiting M dwarfs. M dwarfs spin down slowly and experience elevated levels of high-energy radiation throughout their long pre-main sequence phase, which can last up to 200--300 Myr for mid M dwarfs and up to a Gyr for late M-dwarfs \citep{Baraffe:2015, Godolt:2019}. As a result, understanding the impact of stellar activity on planetary atmospheres is a necessary precursor to identifying habitable conditions for life as described in “Pathways to Habitable Worlds,” one of three priority areas identified in the 2020 Astronomy Decadal Survey \citep{astrodecadal:2020}. Flares occur when magnetic reconnection events in the corona accelerate charged particles along field lines toward the photosphere, heating the plasma. Particles with different energies brake at different depths in the stellar atmosphere and produce different amounts of flux at each wavelength \citep{Klein_Dalla:2017}. The far-UV (FUV; 90--170 nm) and near-UV (NUV; 170--320 nm) emission that drive photochemistry in planetary atmospheres originate in the upper chromosphere and transition region and are observed immediately following an episode of electron beam heating \citep{Fleming:2022}. Optical emission is the most commonly observed component of the flaring process, results from heating in the lower chromosphere, and typically occurs in two phases. Prompt optical emission with an effective temperature of 9,000--16,000 K occurs first, followed by a more gradual period of emission of $\sim$5000 K \citep{Kowalski:2016, Howard:2023}. Here, we define effective temperature ($T_\mathrm{eff}$) or equivalent color temperature as the blackbody temperature that best fits an observed spectral energy distribution or set of broadband observations \citep{Kowalski:2024}.

Transmission spectroscopy observations with JWST are illuminating the role of high-energy stellar radiation for photochemistry in planetary atmospheres. Sulfuric photochemistry has been detected through an SO$_2$ feature at 4.05~$\mu$m in the hydrogen-rich irradiated atmosphere of WASP-39b \citep{Alderson:2023, Rustamkulov:2023, Tsai:2023}. A plethora of Cycle 3 JWST programs have now been selected to explore the impact of photochemical processes on the transmission spectrum (GO 5177, PI: Hu; GO 5191, PI: Ducrot; 5311, 5959, Co-PIs: Feinstein \& Welbanks; GO 5844, PI: Radica; GO 5863, PI: Murray). Although photochemistry is driven by FUV--NUV radiation, the contribution of emission at these wavelengths to the energy budget of flares remains largely unknown. While the Transiting Exoplanet Survey Satellite (TESS; \citealt{Ricker:2014}) has observed more than 10$^6$ flares from 140,000 stars \citep{Gunther:2020, Feinstein:2022a, Yang:2023, Feinstein:2024}, simultaneous UV observations exist for only a few dozen events \citep{MacGregor:2021, Paudel:2021, Jackman:2022, Inoue:2024, Jackman:2024, Paudel:2024}. Of these, only 17 flares have been observed during the flare peak and at sub-minute cadence in each band \citep{Inoue:2024, Jackman:2024, Paudel:2024}. Longer cadences fail to resolve the structure of the flare peak, which evolves on timescales of $\sim$1--30 s \citep{Kowalski:2019, Aizawa:2022, Howard_MacGregor:2022}. Simultaneous UV and optical observations of fifteen flares from six stars have also been obtained using facilities besides \textit{Swift} and TESS \citep{Hawley:1991, Robinson:1993, Osten:2016, Wargelin:2017, Kowalski:2019, Brasseur:2023, Tristan:2023}. The current lack of a statistical sample of simultaneous UV observations prevents us from inferring the photochemical radiation environments of planets from optical flares to within a factor of $\sim$10 \citep{Paudel:2024}. As UV emission from the host star serves as the largest source of uncertainty in photochemical models of small, young planetary atmospheres \citep{Teal:2022} and remains a significant source of uncertainty for larger planets \citep{Hu:2021, Yu:2021}, we require a better understanding of the UV output from stellar flares. Flare emission at UV wavelengths is the largest source of uncertainty because different quiescent spectra result in larger changes to the planetary spectrum than do other model inputs, and because the intensity and spectral shape during flares further increases the uncertainty relative to quiescence.

In the absence of simultaneous observations, the UV emission of optical flares is usually estimated assuming a blackbody spectrum with an $T_\mathrm{eff}$ of $\sim$9000 K (e.g.\ \citealt{Shibayama:2013, Yang:2017, Gunther:2020, Ducrot:2020, doAmaral_2022}). This approach is motivated by the observation that many optical flare spectra are well-described by a spectrum reminiscent of an A0 star or hot blackbody of 9000--10,000 K \citep{Hawley_Fisher:1992, Hawley:2003, Kowalski:2013, Osten_Wolk:2015}. As noted by \citet{Jackman:2023}, applications of the 9000 K blackbody to scaling optical flare energies into the UV involve various degrees of complexity. For example, some UV-optical relations adjust the spectrum of the 9000 K blackbody by including actual UV flare observations \cite[e.g.,][]{Segura:2010, Tilley:2019, Feinstein:2022b, Konings:2022, Louca:2023}. 

However, the accuracy of the 9000 K blackbody scaling is increasingly being called into question as flares are observed in new UV and optical datasets \citep{Jackman:2023, Paudel:2024}. Although the UV and optical emission of some flares are consistent with the 9000 K blackbody \citep{Hawley:1991, Paudel:2021, Inoue:2024}, other events are not. Simultaneous optical and NUV \textit{Hubble} Space Telescope (HST) spectroscopy of two flares from the M4 dwarf GJ 1243 show the NUV emission is underestimated by 2-3$\times$ when scaling from the optical using a 9000 K blackbody due to the presence of Fe II and Mg II lines and ionized Balmer continuum below 364 nm \citep{Kowalski:2019}. Likewise, the FUV emission of a flare from Proxima Cen observed simultaneously by HST and TESS is underestimated by 12.2$\times$ relative to the 9000 K blackbody \citep{MacGregor:2021}. At least 8 flares observed simultaneously with \textit{Swift} and TESS exceed the NUV flux predicted from the 9000 K blackbody by factors of 3--17 \citep{Paudel:2024}, corresponding to $T_\mathrm{eff}$ values of 12,000--26,000 K.

Non-simultaneous UV and optical flare rates also indicate the UV energies of optical flares are underestimated. \citet{Jackman:2023} measure the average flare rates of a sample of 1250 M-dwarfs with GALEX and TESS observations obtained 6--17 years apart. The TESS flare rates are scaled into the FUV and NUV using the 9000 K blackbody and then compared with the observed GALEX NUV flare rates. \citet{Jackman:2023} split the sample into partially and fully convective stars and compute empirical correction factors (ECF) for each sample and flare template. They find the NUV energies derived from the 9000 K blackbody must be scaled by ECFs of 2.7$\pm$0.6 and 6.5$\pm$0.7 to bring the flare rates into alignment with the GALEX observations of the partially and fully convective stars, respectively. Even small underestimates at NUV wavelengths can correspond to much larger underestimates at FUV wavelengths given the assumption that the flare spectrum is described by a blackbody of $T_\mathrm{eff}$. For example, an ECF of 2.7$\pm$0.6 in the NUV corresponds to an ECF of 6.4$\pm$2.8 in the FUV while an ECF of 10$\pm$1 in the NUV corresponds to an ECF in the FUV of 93$\pm$11. Deriving energy budgets from average flare rates obtained in two different wavelength regimes is only possible under the assumption that the processes driving emission in both bands are closely coupled. As a result, \citet{Jackman:2023} state that simultaneous observations are required to confirm any relationship between flare energy and flare $T_\mathrm{eff}$, and to discover under what conditions NUV flares have optical counterparts.

\begin{figure*}
	\centering
	{
		\includegraphics[width=0.98\textwidth]{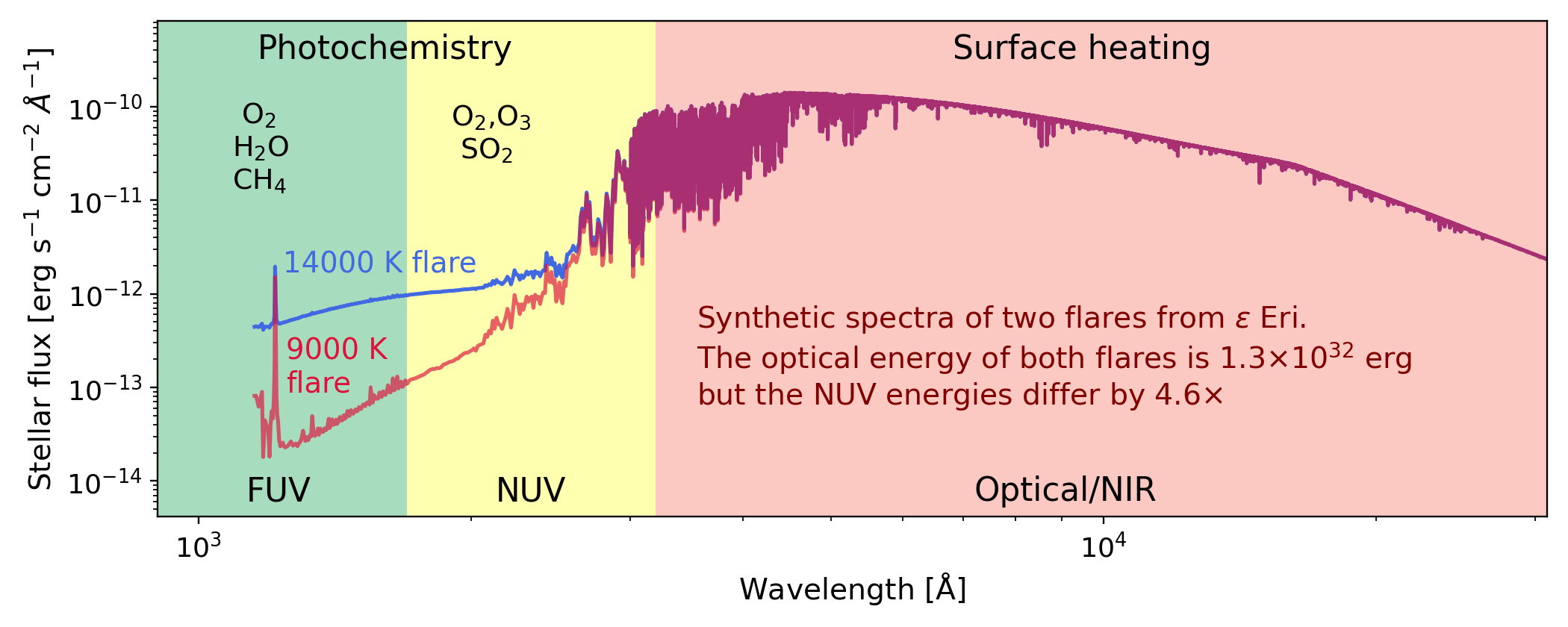}
	}
	
	\caption{Although flare rates and energies are measured in the optical for most stars, the FUV and NUV energies determine the photochemical radiation environment of orbiting planets during these events. Simultaneous observations at UV and optical wavelengths are needed to derive the energy budget as flares with similar emission levels at optical wavelengths can produce NUV fluxes that differ by 2--20$\times$ depending on the $T_\mathrm{eff}$ of the flare, correlating with order-of-magnitude discrepancies at FUV wavelengths. Spectra of $\epsilon$ Eri during the two flares are obtained by superimposing the flare blackbody onto the semi-empirical base spectrum from \citet{Behr:2023}.}
	\label{fig:nuv_optical_sed}
\end{figure*}

Flares also serve as an unpredictable source of stellar contamination in the search for young transiting exoplanets \citep{Newton:2019, Martioli:2021, Gilbert:2022}, biasing the transit parameters or even resulting in false negative detections for particularly active systems \citep{Luger:2017, Gilbert:2021, Mann:2022}. Although flare contamination is worse at shorter wavelengths, flares observed from the young M-dwarf AU Mic during transits of its two warm Neptunes AU Mic b and c in broadband TESS photometry (600--1000 nm) have been observed to alter the ingress and egress shape, reduce the transit depth, and in some cases largely mask the transit itself \citep{Martioli:2021, Gilbert:2021, Mann:2022}. While TESS has been used to detect nearly a dozen new young transiting planets (e.g. \citealt{Benatti:2019, Newton:2019, Plavchan:2020, Mann:2020, Rizzuto:2020, Tofflemire:2021, Zhou:2021, Zhou:2022, Mann:2022, Vach:2022}), the increased stellar activity levels of young stars has lead to substantial losses in the yield of young planets with radii $R<$3R$_\oplus$ and orbital periods $P_\mathrm{orb}>$10 d \citep{Newton:2019, Rizzuto:2020, Zhou:2021}.

Simultaneous monitoring of many young stars at NUV and red optical wavelengths is needed to determine the relative contribution of flares at UV wavelengths and to mitigate flare contamination in searches for young transiting planets. The Early eVolution Explorer (EVE) is a NASA astrophysics Small Explorer concept to investigate the earliest stages in the shared evolutionary pathways of stars and planets. EVE would obtain simultaneous NUV and red optical photometry of the dense core regions of nearby young moving groups and clusters with ages of approximately 5--100 Myr. The mission would observe 12--24 targeted regions for approximately 30 d each and at cadences of 30--60 s. The addition of the simultaneous NUV band compared to missions like \textit{Kepler} or TESS enables the measurement of the NUV-optical relation for hundreds to thousands of flares over the mission lifetime from 10$^4$ stars of different masses and ages, while providing an independent lever arm to identify and remove photometric scatter in transit light curves due to flares. EVE would determine the degree to which the UV radiation environments of young planets are underestimated from optical observations as a function of flare size and shape in order to constrain the impact of photochemistry on sub-Neptune transmission spectra \citep{Konings:2022, Louca:2023} and biosignature false positives for small planets \citep{Harman:2015}. New results from GALEX demonstrate that simultaneous NUV and optical observations with EVE would provide strong constraints on the FUV to NUV ratios of optical flares, which determine the photochemical state of planetary atmospheres. The GALEX observations of the FUV to NUV energy ratios of 182 flares from 158 M dwarfs find that the distribution of FUV/NUV energy ratios of their sample is surprisingly narrow, with a mean and 1$\sigma$ uncertainty of 0.46$\pm$0.23 \citep{Berger:2024}. As a result, a successful NUV-TESS scaling relation would unlock the photochemical radiation environment of the extensive TESS flare archive.

In this work, we demonstrate the feasibility of the EVE flare science objective using new multi-wavelength observations of flares from G 41-14, GJ 674, and EV Lac and new analysis of flares from AP Col and YZ CMi, all obtained simultaneously at 20 s cadence with \textit{Swift} and TESS. In Section \ref{simul_data}, we describe the \textit{Swift} and TESS observations, data reduction, and identification of simultaneous flare detections. In Section \ref{structure_time_lags}, we discuss the methods used to characterize the time lag and energy budget of the flares at each wavelength. In Section \ref{simul_flare_results}, we present population-level results for the consistency of the flare energy budget with a 9000 K blackbody at NUV to optical wavelengths. In Section \ref{transit_detection_efficiency}, we explore improvements to the transit detection efficiency obtained from the addition of the NUV band. In Section \ref{applic_eve}, we present simulations of the number of flares required to discriminate between flare samples with energy budgets consistent with the 9000 K assumption and NUV-luminous flares relative to the 9000 K assumption. We then present a list of candidate clusters and a trade study to ensure EVE can feasibly detect a sufficient number of multi-wavelength flares. In Section \ref{discuss_conclude} we discuss the implications of our work to exoplanet photochemistry and to EVE.

\section{Simultaneous NUV and TESS observations}\label{simul_data}
\begin{figure*}
	\centering
	{
		\includegraphics[width=0.98\textwidth]{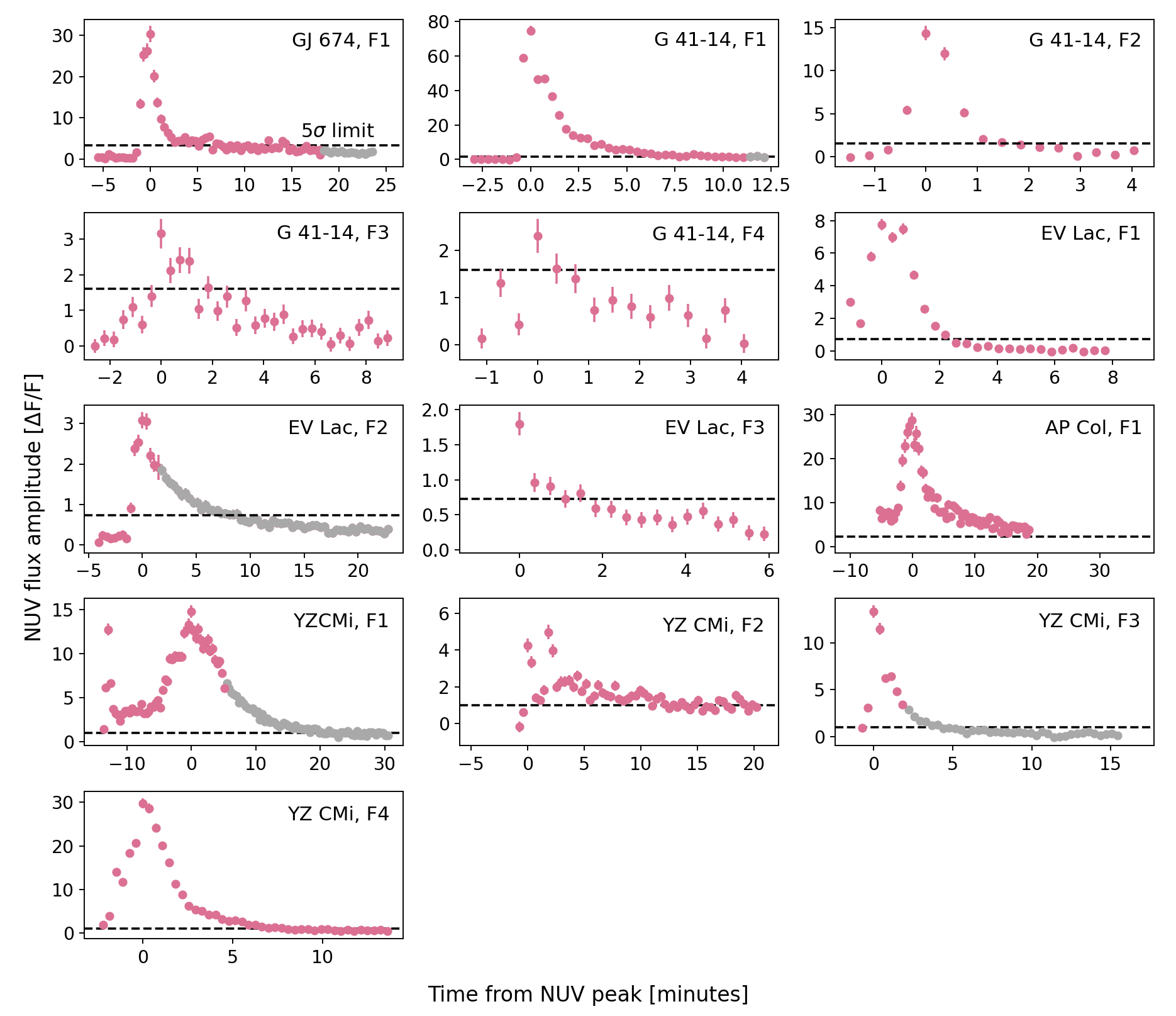}
	}
	
	\caption{NUV 22 s cadence light curves of the 13 flares meeting our detection criteria. Flares are ordered by host star and size. The 5$\sigma$ detection threshold is shown as a black dashed line, and times with simulated light curve data during gaps caused by the \text{Swift} duty cycle are shown in gray. Simulated sections are produced by fitting the \citet{Tovar_Mendoza:2022} flare model to the observed part of the decay phase and varying the points within the expected photometric uncertainties. The flare peak is fully covered by the \textit{Swift} observations during 12 flares and partially covered during one flare (EV Lac F3).}
	\label{fig:nuv_flares_overview}
\end{figure*}

\begin{figure*}
	\centering
	{
		\includegraphics[width=0.98\textwidth]{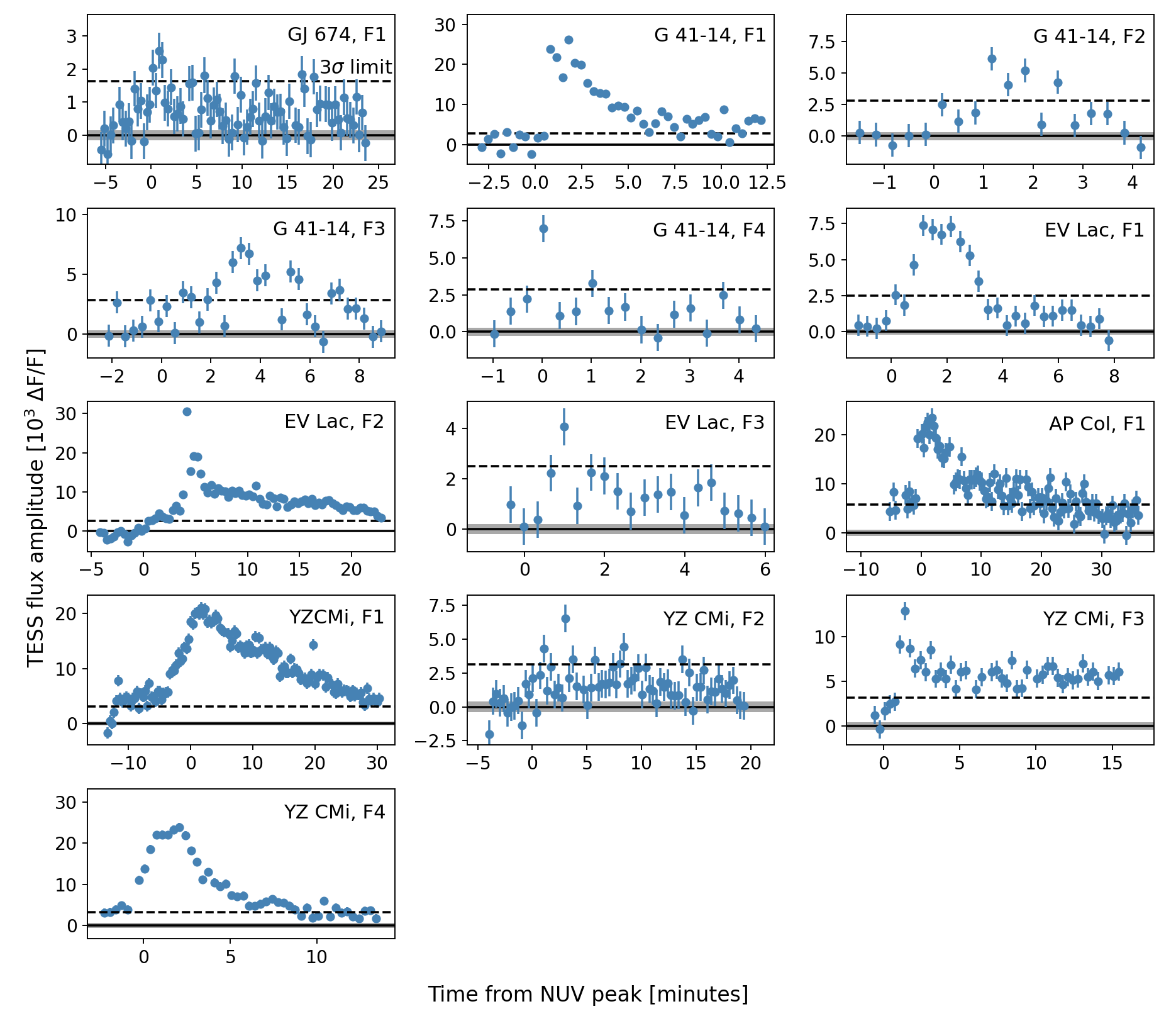}
	}
	
	\caption{Simultaneous 20 s cadence TESS light curves of the 13 flares meeting our detection criteria. The TESS components of the flares are ordered and displayed on the same time axis as in Figure \ref{fig:nuv_flares_overview}. The local flux baseline determined from the average of the nearby out of flare points is shown as a black solid line, and the uncertainty in this baseline value is shown in gray. The 3$\sigma$ detection threshold relative to this value is shown as a black dashed line.}
	\label{fig:tess_flares_overview}
\end{figure*}

\begin{table}
\centering
\caption{Target Sample and Simultaneous TESS and \textit{Swift} Observations}
\begin{tabular}{ccccccccccccc}
\hline
\hline
\hline
Star & SpT & Age & $d$  & $T$ \\
     &     & Myr & pc & \\
\hline
GJ 674 & M3V & 100--1000 & 4.553$\pm$0.001 & 7.158$\pm$0.008  \\
G 41-14 & M3.5V & 100--600 & 6.804$\pm$0.07 & 8.206$\pm$0.014 \\
EV Lac & M4.0V & 125--800 & 5.0516$\pm$0.0006 & 7.730$\pm$0.007 \\
AP Col & M5 & 12--50 & 8.666$\pm$0.002 & 9.655$\pm$0.008 \\
YZ CMi & M4.5V & $\sim$200 & 5.989$\pm$0.001 & 8.337$\pm$0.007 \\
 & &  & & \\
\hline
\hline
Star & $t_\mathrm{obs,~NUV}$ & $n_\mathrm{visits,~NUV}$ & $t_\mathrm{start,NUV}$ & $t_\mathrm{end,NUV}$ \\
     & ks                    &                          & UTC                    & UTC \\
\hline
GJ 674 &  37.9 & 5 & 2023-06-17 01:11:38 & 2023-06-28 04:08:17.2 \\
G 41-14 & 8.52 & 3 & 2023-01-26 02:24:36 & 2023-01-29 03:05:47 \\
EV Lac & 10.4 & 10 & 2022-10-24 12:02:34 & 2022-10-27 15:10:40 \\
AP Col & 17.82 & 10 & 2020-12-06 00:02:34 & 2020-12-07 16:11:56 \\
YZ CMi & 17.38 & 2 & 2021-01-23 04:32:34 & 2021-01-23 11:27:57 \\
 & &  & & \\
\hline
\hline
Star & $t_\mathrm{obs,~TESS}$ & $n_\mathrm{sectors,~TESS}$ & $t_\mathrm{start,TESS}$ & $t_\mathrm{end,TESS}$ & $Q_0$ \\
     & d                      &                            & UTC                     & UTC & erg s$^{-1}$ \\
\hline
GJ 674 & 26.9 & 66 & 2023-06-02 04:17:25 & 2023-06-30 22:23:03 & 10$^{31.137}$ \\
G 41-14 & 24.5 & 61 & 2023-01-18 07:17:45 & 2023-02-12 17:33:41 & 10$^{31.066}$ \\
EV Lac & 25.0 & 57 & 2022-09-30 20:35:42 & 2022-10-29 14:50:33 & 10$^{30.998}$ \\
AP Col & 25.0 & 32 & 2020-11-20 17:16:49 & 2020-12-16 17:39:48 & 10$^{30.697}$ \\
YZ CMi & 24.1 & 34 & 2021-01-14 06:20:38 & 2021-02-08 13:42:44 & 10$^{30.903}$ \\
\hline
\hline
\hline
\end{tabular}
\label{table:summary_sample_and_obs}
{\newline\newline \textbf{Notes.} Description of the observations used in this work. Columns in the top panel are the star, spectral classification, age estimate, stellar distance, and TESS magnitude. We obtain age estimates of GJ 674 from \citet{Bonfils:2007}, EV Lac from \citet{Paudel:2021}, AP Col from \citet{Riedel:2011}, and age estimate for G 41-14 and YZ CMi using a rotation-age relationship from \citet{Engle:2023} as described in the main text. TESS magnitudes and stellar distances are from \citet{Bailer_Jones2018} and \citet{Gaia:2018, Gaia:2021}, respectively.}
\vspace{0.05cm}
\end{table}

\subsection{Sample selection}\label{sec:target_sample}
We analyze simultaneous photometry of the mid M flare stars GJ 674, G 41-14, EV Lac, AP Col, and YZ CMi at NUV and red optical wavelengths with \textit{Swift} and TESS. These stars are among the nearest, brightest, and most flare-active M-dwarfs in the TESS field. Each of these stars was originally observed by \textit{Swift} and TESS during larger coordinated flare monitoring programs. \textit{Swift} observations of G 41-14 (ToO 18362; PI: Howard) and GJ 674 (ToO 18954; PI: Howard) were obtained in coordination with ALMA and TESS as part of a campaign to understand the connection between millimeter and shorter wavelength flare emission in a sample of nearby M-dwarfs (e.g., \citealt{MacGregor:2021, Howard:2022, Burton2021AAS}). \textit{Swift} observations of EV Lac (ToO 17987; PI: Inoue) were obtained in coordination with NICER, Swift, TESS, and the 2 m Nayuta telescope at the Nishi-Harima Astronomical Observatory to understand the spectral model of flares from X-ray to red optical wavelengths \citep{Inoue:2024}. AP Col and YZ CMi were observed as part of a comprehensive multi-wavelength survey of nearby M dwarf flare stars at X-ray to optical wavelengths using TESS, Kepler/K2, \textit{Swift}, and HST \citep{Paudel:2024}. The 20 s TESS data and simultaneous \textit{Swift} data was obtained for the TESS GO program 3273 (PI: Vega). These observations provide a representative dataset in which to assess the range of multi-wavelength flare properties and contamination that are likely to be present in long-term observations by EVE.

In addition to the properties of each star listed in Table \ref{table:summary_sample_and_obs}, we summarize the literature on stellar flares from our sample. GJ 674 is known to flare at FUV wavelengths \citep{Froning:2019} and host a non-transiting 11M$_\oplus$ planet on a 4.69 d orbit \citep{Bonfils:2007}. G 41-14 is a spectroscopic binary with an orbital period of 7.555$\pm$0.002 d \citep{Delfosse:1999} and rotational variability present at the 1\% level in its 20 s TESS light curve. We estimate its age to be 100--600 Myr based on the period $P_\mathrm{rot}$=0.799$\pm$0.011 d present in the TESS light curve, the M2.5--6.5 rotation-age relation of \citet{Engle:2023}, and a binarity correction factor of 3.3 on the age upper limit inferred for a single M3.5 dwarf with $P_\mathrm{rot}<$5 d \citep{Fleming:2019}. EV Lac is a benchmark flare star with a long history of flare observations in various wavelength regimes (e.g. \citealt{Pettersen:1976, Zhilyaev:1998, Osten:2010, Paudel:2021}). AP Col is a pre-main sequence star that is frequently observed to flare in the TESS bandpass \citep{Medina:2020, Howard_MacGregor:2022}. YZ CMi is a well-known flare star and has been observed to flare at X-ray to radio wavelengths (e.g. \citealt{Lacy:1976, Doyle:1988, Kowalski:2010, Maehara:2021, Notsu:2024}). The 2.77 d \citep{Maehara:2021} rotation period of YZ CMi suggests an estimated age of $\sim$200 Myr \citep{Engle:2023}. 

\subsection{NUV Observations with \textit{Swift}}\label{sec:uvot_obs}
Monitoring of each star was performed with the \textit{Swift} Ultraviolet/Optical Telescope (UVOT; \citealt{Roming:2005}). The observations and monitoring time are given in the second part of Table \ref{table:summary_sample_and_obs}. G 41-14, GJ 674, AP Col, and YZ CMi were each observed in the UVM2 filter (1990--2500 $\AA$), while EV Lac was observed in the UVW2 filter (1600--2260 $\AA$).

We reduced the unscreened level one event files by following the recipe given in the \textit{Swift} UVOT Software Guide \citep{Immler:2008}. First, we repopulated the coordinate columns in the unscreened level one event files with the FTOOL \texttt{coordinator} tool in \texttt{heasoft 6.29}. We also updated the stellar positions from J2000 to J2022 values due to the high proper motions of each star during the \texttt{coordinator} step. Next, we created screened level two event files with the \texttt{uvotscreen} tool. As described in \citet{Inoue:2024}, the quality flags of valid events in some observations have values greater than zero and can result in empty level 2 event files if the standard \texttt{uvotscreen} settings are used. We decided to include all events regardless of quality flag status in affected observations after performing a visual inspection of the point spread function at the target coordinates in the images to ensure the absence of cosmic ray strikes or other anomalies.

We created source and background region files for each level two event file in \texttt{ds9} as described in \citet{Paudel:2021}, where source regions are defined by an aperture of 5" centered on the proper-motion corrected target coordinates and background regions are defined by a 30" aperture centered on a blank part of the image. The image quality sometimes necessitated a slightly larger source radius of $\sim$7". We extracted calibrated light curves for the source and background regions using the \texttt{uvotevtlc} tool at a uniform 22.066 s cadence to observe flare substructure at a comparable time-resolution to the TESS data. Only cadences that are multiples of the UVOT minimum time resolution of 11.033 ms are allowed by the \texttt{uvotevtlc} architecture. We access calibration data via a file transfer protocol connection to the Goddard Space Flight Center Calibration Database (CALDB). Pile-up mitigation is implemented by \texttt{uvotevtlc} using CALDB \citep{Paudel:2021}. Finally, the light curves are barycenter-corrected using the \texttt{barycorr} FTOOL to place them on the same time axis as the TESS data.

\subsection{Optical TESS Observations}\label{sec:tess_obs}
TESS observed GJ 674 (TIC 218263393), G 41-14 (TIC 283784587), EV Lac (TIC 154101678), AP Col (TIC 160329609), and YZ CMi (TIC 266744225) at 20 s cadence as described in the third part of Table \ref{table:summary_sample_and_obs}. The TESS observations completely covered the UVOT observations for a total of 27.1 hr of simultaneous monitoring across all five stars. We note not all overlapping coverage is usable due to times with low NUV count rates.

TESS observes the entire sky to enable the search for exoplanet transits and other sources of variability, split into $\sim$28 d sectors \citep{Ricker:2014}. Each 96$\times$24 degree TESS sector is observed continuously using four 10.5 cm telescopes in a red optical (600-1000 nm) bandpass at 21$\arcsec$ pixel$^{-1}$. Downlink gaps of $\sim$1 d occur once per orbit, approximately every 13.7 d. Fast (20 s) TESS light curves of our target stars were downloaded from the Mikulski Archive for Space Telescopes \cite[MAST;][]{10.17909/t9-st5g-3177} \footnote{https://mast.stsci.edu. The specific observations analyzed can be accessed via \dataset{10.17909/t9-st5g-3177}}. We downloaded Simple Aperture Photometry (SAP) instead of Pre-search Data Conditioning (PDC) light curves in order to apply our own detrending routine and to prevent the suppression of flare flux in the PDC data.

We detrend stellar variability in the light curves following the process described for the 20 s flare sample in \citet{Howard_MacGregor:2022}. Briefly summarizing, a Savitzky–Golay (SG) filter is applied to all points in the light curve within 2.5$\sigma$ of the average flux to identify slowly-evolving variability. The SG model is subtracted from the light curve and flare candidates are identified in the residual light curve as excursions that reach $\geq$3$\sigma$ above the photometric noise. Candidate flares are vetted by eye, ensuring the start and stop times include the entirety of the rise and decay phases and any flares missed during the automated detection are included in the event list. A second SG filter is then applied to all points in the original light curve outside of the confirmed flare events and subtracted to produce a final detrended light curve. The SG window length is determined by the cadence and stellar rotation period, where the 20 s cadence window length is 303 points for $P_\mathrm{rot} <$ 1 d, 453 for 1 $\leq P_\mathrm{rot} <$ 2 d, and 603 points for $P_\mathrm{rot} \geq$ 2 d. The 1$\sigma$ flux errors of the detrended and normalized fluxes of GJ 674, G 41-14, EV Lac, AP Col, and YZ CMi during the out-of-flare times are $\pm$0.0006, $\pm$0.0012, $\pm$0.0008, $\pm$0.0021, and $\pm$0.0011, respectively.

\subsection{Identifying simultaneous events}\label{sec:identifying_simul_flares}
Since the flare response is strongest in the \textit{Swift}/NUV, we first identify simultaneous flares in the \textit{Swift} light curves to ensure the inclusion of small events. Counterparts to each \textit{Swift} detection are then confirmed in the TESS light curves. We identify flares in the \textit{Swift} light curves as events containing at least two adjacent 5$\sigma$ excursions above the median out-of-flare flux. The adjacent 5$\sigma$ requirement is imposed because smaller flux increases in the \textit{Swift} light curves usually do not have clear TESS counterparts and are therefore not useful for the purposes of our study. We find quiescent count rates of 0.49$\pm$0.04 cps for GJ 674, 1.26$\pm$0.06 for G 41-14 cps, 0.50$\pm$0.12 cps for AP Col, and 1.98$\pm$0.27 cps for YZ CMi in the UVM2 band, and a quiescent count rate of 6.05$\pm$0.11 for EV Lac in the UVW2 band. Only 13 flare peaks are detected in the \textit{Swift} light curves given these criteria, one event from GJ 674, four from G 41-14, three from EV Lac, one from AP Col, and four from YZ CMi. The NUV light curves of these flares are shown in Figure \ref{fig:nuv_flares_overview}. The NUV start and stop times, energies, and detection significances are given in Tables \ref{table:flare_timings} and \ref{table:flare_energies}.

We assume a 9000 K blackbody spectrum and neutral hydrogen column density of 3$\times$10$^{18}$ cm$^{-2}$ to convert the \textit{Swift} count rates produced by \texttt{uvotevtlc} into flux values in units of erg s$^{-1}$ cm$^{-2}$ integrated over the UVM2 wavelength range of 1990--2520 $\AA$ using WebPIMMS version 4.12. Upper and lower systematic uncertainties on the conversion are defined by blackbody temperatures of 8000 and 16,000 K, which are representative of the range of temperatures usually observed for the hot blackbody component that dominates at NUV wavelengths \citep{Kowalski:2013, Kowalski:2016}. The systematic uncertainty impacts the UVM2 counts to flux conversion by 1.6\% and 3.2\% for the lower and upper temperatures, respectively. The larger uncertainties for the lower and upper temperatures of 1.6\% and 18.1\% for the UVW2 counts to flux conversion are due to the difference in wavelength coverage of the two bands. In both cases, the difference in total energy between the 8000 and 16,000 K scenarios remains small relative to the power-law scaling relationships explored in this work. The luminosity at the stellar surface is computed in units of erg s$^{-1}$ using the stellar distance. Flare energies are measured in erg between the start and stop times of each \textit{Swift} flare from Table \ref{table:flare_timings}.

The TESS counterparts are often weaker than the larger flares picked up in the automated detrending process. The ``quiescent" flux baseline of very active stars appear consistent with constant low-level flaring, which can be seen in Fig. 2 of \citet{Gilbert:2021} for the TESS light curve of Proxima Cen after subtraction of the larger flares. In that work, the TESS fluxes below the median are consistent with that expected for a 0.43 ppt photometric error at 2 min cadence for a T=7.36 star, while several dozen flux increases of $\sim$1.5 ppt are seen above the median in each sector. The low-level activity decreases the apparent significance of the TESS counterparts by biasing the out-of-flare baseline value upwards. As a result, determining the significance of small flares relative to the local flux baseline is challenging. We therefore visually inspect the light curves within $\sim$30 min of each TESS counterpart in order to create a mask excluding all times that appear consistent with a fast-rise, exponential decay flare morphology. TESS counterparts are then identified as two consecutive 3$\sigma$ points relative to the mean flux of the remaining unmasked points. This detection criterion is chosen because it recovers all of the \textit{Swift} flares without also including further low-level variability. The TESS light curves of these flares are shown in Figure \ref{fig:tess_flares_overview}. We record the start and stop times of the FWHM duration and measure the energy emitted in this period as described in \citet{Howard:2020}. The quiescent luminosity in the TESS band is measured in erg s$^{-1}$ using the TESS mag and stellar distance, and the equivalent duration is measured in s from the fractional flux relative to the local out-of-flare flux. The TESS FWHM start and stop times, quiescent luminosities, energies, and detection significances are given in Tables \ref{table:summary_sample_and_obs}, \ref{table:flare_timings}, and \ref{table:flare_energies}.

Parts of the decay phases of Flare 1 from GJ 674, Flare 1 from G41-14, and Flare 3 from EV Lac occurred outside the \textit{Swift} window. We fit the \citet{Tovar_Mendoza:2022} flare model to the parts of the flare decay observed within the \textit{Swift} window and extrapolate the likely drop in flux outside the window. We vary the simulated fluxes within their photometric uncertainties to ensure maximum accuracy. Since we have TESS observations during the missing \textit{Swift} times, the simulated flare tail is only extrapolated to times where no new flare peaks are seen in the TESS light curve. The simulated parts of the \textit{Swift} flare light curves are highlighted in gray in Figure \ref{fig:nuv_flares_overview}. Times with simulated data are shown in gray, sigma detection limits in black, baseline in black with uncertainty in gray.

\begin{figure*}
	\centering
	{
		\includegraphics[width=0.98\textwidth]{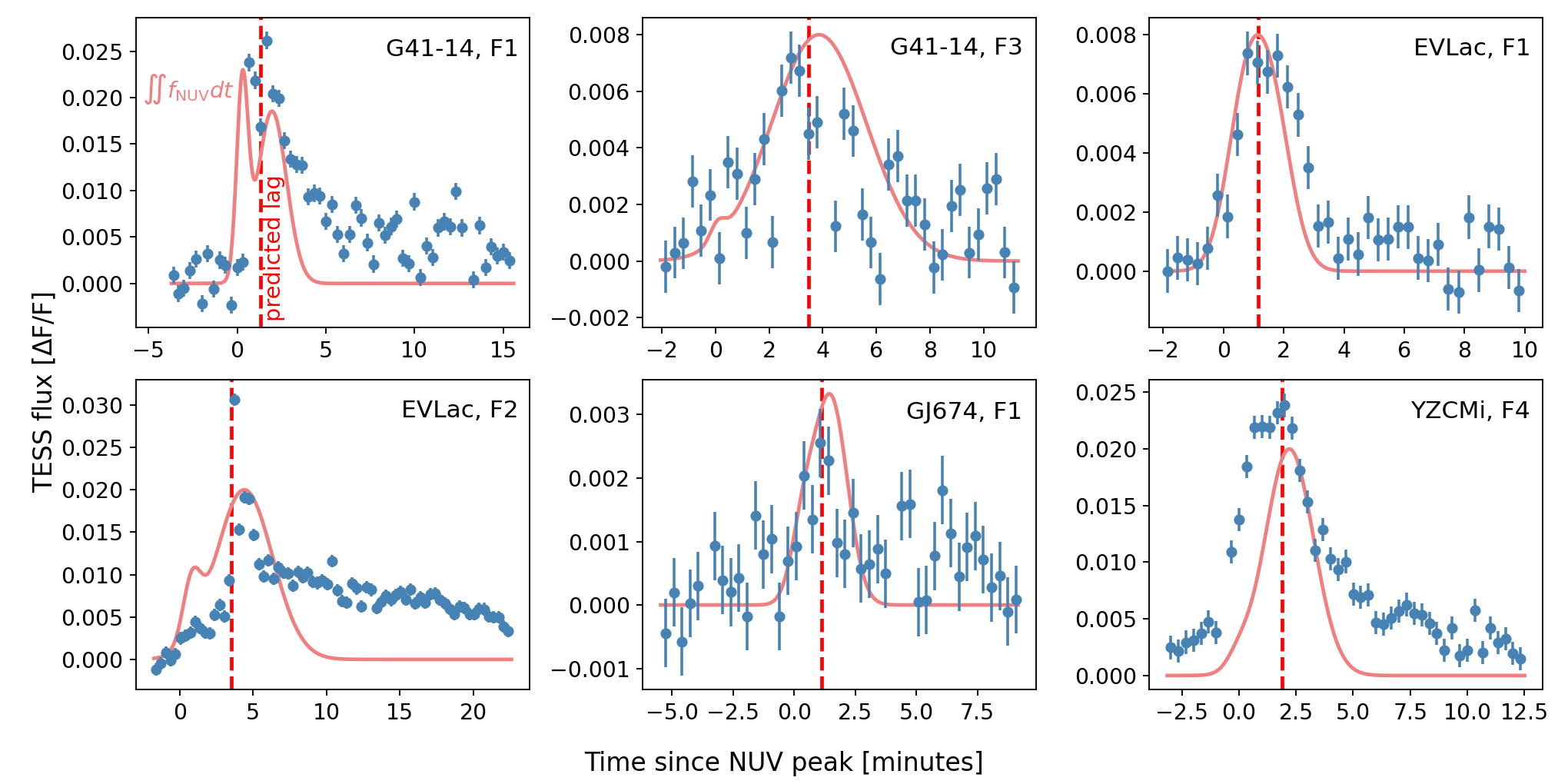}
	}
	
	\caption{Predicted times for representative TESS band light curves relative to the times of the NUV peaks (which occur at $t$=0 on the x-axis and are not otherwise shown). A one- or two-component Gaussian is fit to the peak of the NUV light curve and the double time integral is computed, shown for reference here in salmon. The predicted time of the TESS peak is then given as the time compared to the NUV peak.}
	\label{fig:time_lag_explainer}
\end{figure*}

\begin{figure*}
	\centering
	{
		\includegraphics[width=0.98\textwidth]{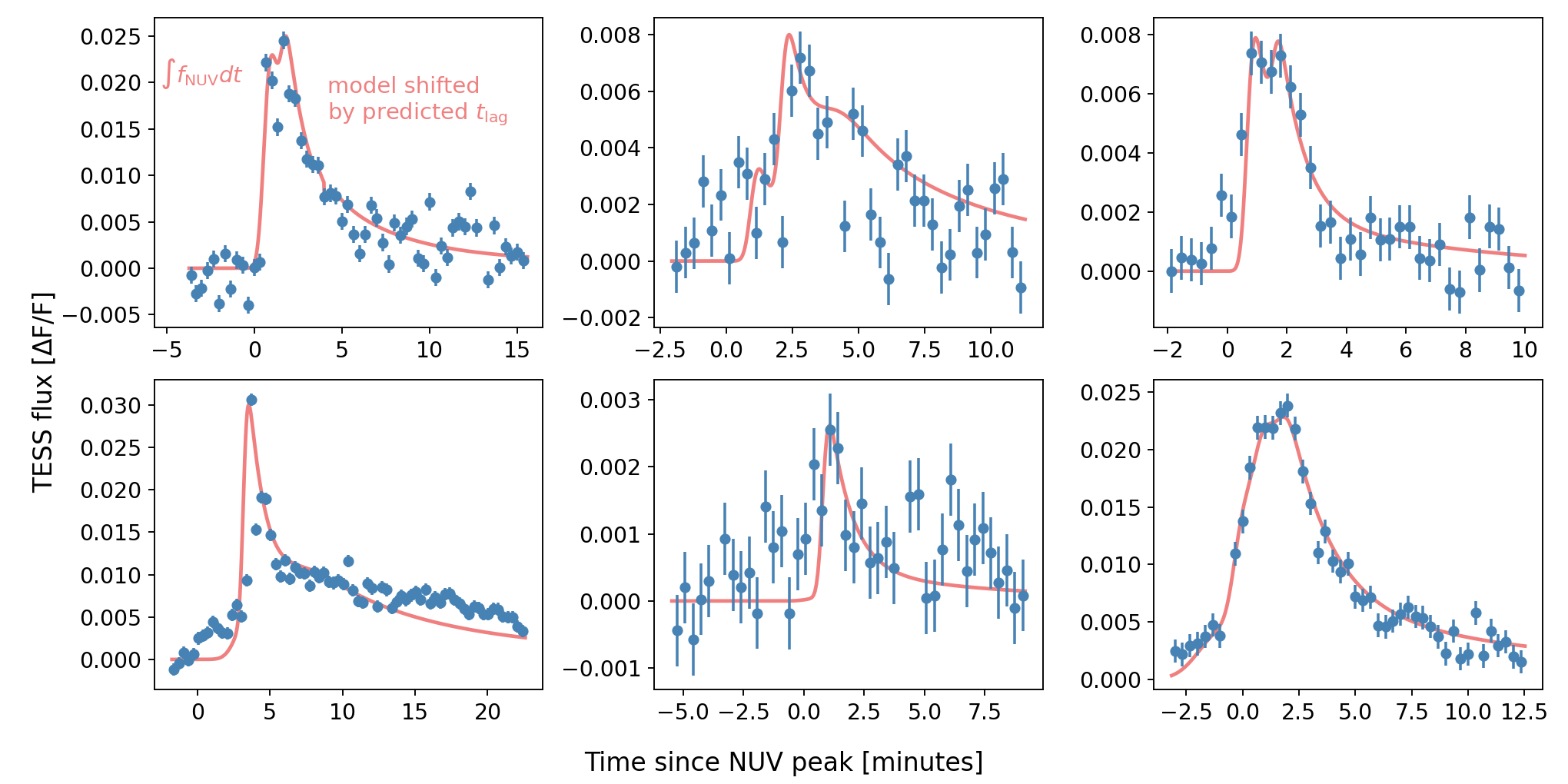}
	}
	
	\caption{Predicted shapes for representative TESS band light curves based on the single time integral of the model fit to the corresponding NUV light curve. The predicted shapes are a superposition of the separate component peaks and the resulting complex flare model's time axis is adjusted by the time lag shown in Fig. \ref{fig:time_lag_explainer}. The flare amplitude of each component is left free to fit to the TESS flux variability and is not a prediction.}
	\label{fig:flare_shape_explainer}
\end{figure*}

\section{Predicting the timing and structure of the TESS flare light curve}\label{structure_time_lags}
Flares are known to reach peak brightness at UV wavelengths prior to optical wavelengths \citep{MacGregor:2021, Inoue:2024}. This time lag between the NUV and optical peak has been observed in a handful of TESS flares, but no systematic exploration has been made to predict the amount of lag. The ability to predict the time and structure of the TESS flare peak solely from the NUV light curve would improve the detection efficiency of transit searches for young, active stars by mitigating correlated stellar noise. The independent source of information on the timing and shape of the TESS band flares provides a lever arm for identifying and removing low-level variability due to small flares and false positive transit events caused by flux dips between the successive peaks of large flares.

Our approach to predicting the time lag and structure of the TESS flare components draws inspiration from the Neupert Effect \citep{Neupert:1968} as seen on the Sun. In the same way that the UV peak occurs prior to the soft x-ray (SXR) peak as a result of the slower time scale of coronal heating relative to prompt UV emission \citep{Tristan:2023}, flare emission at red optical wavelengths results from heating processes that also follow the initial process of particle injection \citep{Kowalski:2016, MacGregor:2021, Howard:2023}. In the Neupert Effect, the time integral of the NUV light curve is predictive of the shape and peak time of the SXR light curve. We extend this time integration approach to predict the time lag and shape of the TESS light curve resulting from chromospheric heating as described below \citep{Neupert:1968}.

Complex or multi-peaked flares occur when several individual flare peaks are super-imposed in the light curve \citep{Howard_MacGregor:2022}. When multiple peaks are present in the NUV light curve of a complex flare, we obtain the model parameters for each individual component separately using joint fits or by subtraction of fits to earlier peaks prior to fitting the subsequent ones. For example, the NUV flare peaks are often composed of two dominant components. The set of time-integrated model parameters for each individual peak are determined from the corresponding set of NUV model parameters using the custom flare integration routine described in \S \ref{custom_intg_routine}.

\subsection{Prediction of the time lag between NUV and TESS band flare peaks}\label{predict_tess_lag_method}
The Neupert Effect applies to coronal heating traced by soft X-rays, while the TESS band traces chromospheric heating. As a result, we do not make an a priori assumption that a single integral will predict both the TESS light curve shape and time lag. We therefore test both single and double integrals, finding the double integral better predicts the time lag for 10 of 13 flares with improvements of 37$\substack{+65 \\ -12}$ s. Although the double time integral of both the empirical flare template and Gaussian light curve model each correlate with the timing of the TESS flare peak, we find the Gaussian model works best for predicting the time lag. The Gaussian model does a better job of describing the rounded peaks observed in the NUV, is more robust against degenerate fits to the NUV peak and over-fitting, and emphasizes the shape of the peak rather than the long decay tail. Prediction of the time lag depends primarily upon the peak times of emission and becomes increasingly inaccurate as more of the decay tail is included in the fit. In fact, the centroid time of the flare peak appears to characterize the cumulative effect of flare heating on the observed flux in each band. We therefore adopt the following approach when predicting the time lag for each flare:
\begin{enumerate}
    \item Fit single and double Gaussian models to the flare peak in the NUV light curve. Only the two largest peaks in each event are needed for the time lag prediction, since not all component peaks are equally important to the prediction process
    \item Determine whether the single or double Gaussian better describes the NUV peak shape. To do this, we subtract each model from the NUV flux within the flare FWHM and define $\mathcal{L}_\mathrm{G2/G1}$ as the ratio of the residual sum of squares of each model (RSS) within the FWHM. $\mathcal{L}_\mathrm{G2/G1}$ is defined such that values greater than one correspond to smaller RSS values for the double Gaussian and favor the two component fit. Visual inspection finds that a portion of the decay tail is included within the FWHM times for several flares. We mask these points from the $\mathcal{L}_{2/1}$ calculation for the affected events.
    \item The flux values during the FWHM times are varied within the flux errors 1000 times to obtain a $\mathcal{L}_\mathrm{G2/G1}$ distribution. The robustness of the $\mathcal{L}_\mathrm{G2/G1}$ measurement is determined based on the fraction of trials in which the double Gaussian model is preferred.
    \item The single or double Gaussian fit is selected, and each component of that fit is passed through our integration routine twice. The second time integrals of each component are then added together to create a final model of the integrated flux. An exception is made in the case of AP Col F1, where the one-component fit results from the FWHM being broader than the peak separation, so the two-peak fit is used instead.
    \item The weighted average of the times of the two largest components is computed using the area under the two curves and adopted as the predicted time of the TESS peak, $t_\mathrm{pred}$.
\end{enumerate}
This procedure is illustrated for six flares in our sample in Figure \ref{fig:time_lag_explainer}. We report a 3$\sigma$-clipped mean and standard deviation of 36$\pm$30\% on the accuracy of the time lag prediction relative to the actual centroid time of the TESS peak. The predicted TESS peak time is compared with the actual peak time to assess the accuracy of the lag prediction. The actual TESS peak time is computed as the centroid within the FWHM of the flare light curve for most flares. The centroid times of several flares (G 41-14 F3, EV Lac F1, AP Col F1 and YZ CMi F1) are biased toward the decay phase by 20--40 s due to complex flaring or the presence of part of the decay tail within the FWHM. For these flares, we compute the actual TESS peak time as the central time of a Gaussian fit to the first dominant peak in the TESS light curve.

Ambiguity exists about which points in the NUV light curve should be included in the Gaussian fit to the flare peak, especially as the rapid phase transitions into the decay phase. As a result, unintentionally choosing the times around the peak that give the best prediction could introduce a bias to the reported accuracy. Therefore, we perform 1000 Monte Carlo (MC) tests for each flare to assess the effect of the time range and fit parameters on the fragility of the lag prediction. In each trial, the NUV fluxes are varied within their errors and the time range used in the original Gaussian fit is randomly shifted forward or backward by one time point. For the flare sample, the mean offset of the predicted time from the accepted value is 20.3 s.

\subsection{Prediction of the TESS flare light curve morphology}\label{predict_tess_shape_method}
Each peak in the TESS flare is fit using the empirical flare template in order to decompose the flare into its constituent events, with each component described by a set of $t_{NUV}$, $\sigma_\mathrm{NUV}$, and $A_\mathrm{NUV}$ values. Here, $t_{NUV}$ is the NUV peak time, $\sigma_\mathrm{NUV}$ is the width of the NUV flare component, and $A_\mathrm{NUV}$ is the fractional flux amplitude of the NUV component. These parameters are then passed to our integration routine in order to obtain the set of empirical flare template parameters that best describe the time integrated flare, $t_\mathrm{TESS}$, $\sigma_\mathrm{TESS}$, and $A_\mathrm{TESS}$. The components are superimposed to create a complex flare light curve and the $t_\mathrm{TESS}$ parameters are shifted by $t_\mathrm{pred}$ - $t_\mathrm{model}$, where $t_\mathrm{model}$ is the centroid of the double integral flare models. Finally, the $A_\mathrm{TESS}$ values are scaled to match the flux values of the actual TESS light curve at the corresponding times. Only the amplitude-scaling process requires information obtained from the optical data via the $A_\mathrm{TESS}$ free parameter, lowering the risk of removing potential transits along with the stellar variability. When present, we note that any degeneracy in the choice of fit parameters to the NUV light curves of complex flares can worsen or improve the model prediction. This procedure is illustrated in Figure \ref{fig:flare_shape_explainer}.

We quantify the observation that a single integral model best explains the shape of the TESS flare light curve by comparing the RSS values of the single time integral model with a double time integral model and no-integration model (i.e., the NUV flare model). For the double and no-integration models, the time-lag and amplitudes of each component peak are scaled to best fit the TESS light curve shape. Next, the RSS ratio for each model is computed for the full duration of each flare to determine how often the single time integral is preferred. We find that flares with low SNR values do not strongly favor one model over the other and therefore limit our analysis to the 7 largest flares with an SNR of 5$\sigma$ at half the maximum flux amplitude. These 7 large flares are G41-14 F1, EV Lac F1, EV Lac F2, AP Col F1 , YZ CMi F1, YZ CMi F3, and YZ CMi F4. The RSS ratio distribution of these flares favors the single-integration model with a mean and mean uncertainty of 6.8$\pm$4.1 and 83$\pm$62 for the no-integration and second-integral models, respectively. The SNR cut ensures that the shape of both the flare peak and decay phase are clearly defined. Since the sample is small, a 1$\sigma$ binomial Confidence Interval (CI) is computed to determine for what fraction of the flares the single integral model is preferred. We find the single integral model is preferred over the no-integration model for 71$\substack{+14 \\ -19}$\% when measured for the full flare duration. The single integral model is likewise preferred for 86$\substack{+9 \\ -18}$\% of the sample relative to the double-integration model.

\subsection{Custom routine for obtaining time integrals of NUV flare models}\label{custom_intg_routine}
Photometric noise and flare substructure present in both the NUV and TESS bands complicate the integration process. First, we find that fitting a flare model to the NUV light curve points and integrating the model is more robust to noise than direct numerical integration of the individual points themselves. We adopt the \citet{Tovar_Mendoza:2022} empirical flare template as our primary model, but also try fitting a Gaussian model to the large-scale structure of the NUV flare peak. Second, we find it is simplest to work with an integration process whose inputs and outputs are each given in terms of the flare model fit parameters of peak time $t_\mathrm{band}$, width $\sigma_\mathrm{band}$, and flux amplitude $A_\mathrm{band}$. To accomplish this, we write a routine that identifies the parameters of a model whose derivative matches the observed UV flare shape. The routine tests a range of parameters for either the \citet{Tovar_Mendoza:2022} empirical flare template or the Gaussian model, computes the time derivative of the tested models, and identifies the set of output parameters $t_\mathrm{TESS}$, $\sigma_\mathrm{TESS}$, and $A_\mathrm{TESS}$ for the model whose derivative provides the optimal fit to the input model with parameters $t_{NUV}$, $\sigma_\mathrm{NUV}$, and flux amplitude $A_\mathrm{NUV}$. Requiring the integration inputs and outputs to each be in terms of a simple flare template limits the analysis to well-motivated fits and makes the process more scalable to transit searches across thousands of stars. What might otherwise be a complex and computationally-intensive fitting process is reduced into $N_\mathrm{peaks}$ separate applications of the integration routine in order to obtain the set of $t_\mathrm{i}$, $\sigma_\mathrm{i}$, and $A_\mathrm{i}$ model parameters for each of the $N_\mathrm{peaks}$ peaks. In a similar way, predicting the effect of multiple stellar flares across the entire optical light curve of a target star is reduced to $N_\mathrm{flare}$ applications of the routine for $N_\mathrm{flare}$ flares.
\begin{figure*}
	\centering
	{
		\includegraphics[width=0.88\textwidth]{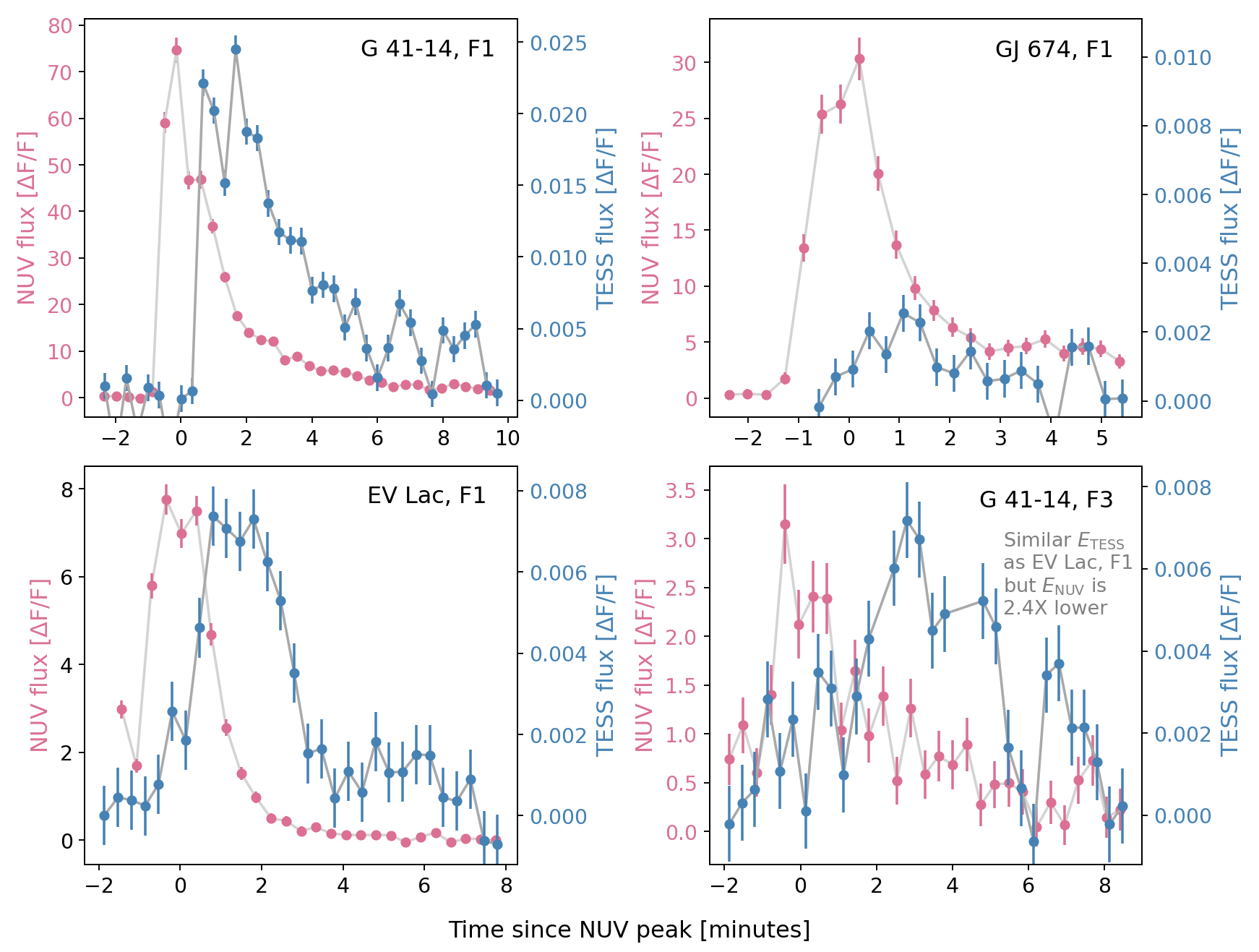}
	}
	
	\caption{Top panels: Simultaneous 20 s cadence light curves of the G 41-14 F1 and GJ 674 F1 events. Although the NUV light curves of both flares are clearly visible, the significance of the TESS light curve of GJ 674 F1 appears much weaker than that of G 41-14 and would have been missed at a lower cadence. The y-axis of the GJ 674 plot has been scaled by the ratio of the TESS to NUV flux amplitude for the G 41-14 flare in order to show the lower TESS to NUV flux ratio of the GJ 674 flare. Bottom panels: Same as the top panels, but for the EV Lac F1 and G 41-14 F3 events. Although the TESS band energies of the flares are comparable, the NUV energy of the EV Lac flare is 2.4$\times$ higher than that of the G 41-14 flare.}
	\label{fig:twenty_sec_results}
\end{figure*}

\subsection{Reversibility of NUV and TESS prediction processes?}\label{reverse_procedure}
Inferring the time lag and shape of the NUV light curve from the TESS light curve is more difficult than for the reverse scenario. Two factors are primarily responsible for this, which are the (1) greater stellar contrast of M dwarf flares at NUV than TESS wavelengths and (2) greater impulsiveness/temporal separation of the component peaks of complex flares at NUV wavelengths. The S/N of even the largest optical counterparts of the NUV flares in our sample is lower than for the NUV components due to the reduced stellar contrast of flares at optical wavelengths, while overlapping and unresolved optical peaks result in uncertainty about the correct number and shape of the component fits. We assess the relative performance of the forward (TESS) and reverse (UVM2) light curve predictions using the same seven high-S/N flares listed above. The TESS light curves are fit using the flare template, and the first and second derivatives for each model component are taken. Two differences from the forward procedure are elimination of the use of the Gaussian model due to the less impulsive shape of the TESS light curve, and the use of only the dominant flare component when estimating the reverse lag. We find the 63$\substack{+185 \\ -45}$\% errors on the reverse lag prediction to be indistinguishable from those previously measured for the forward process (36\%). We measure reduced $\chi^2_\mathrm{red}$ values for the forward and reverse scenarios and find lower $\chi^2_\mathrm{red}$ values for four of seven events. These four favor the forward procedure by $\mathcal{L}_\mathrm{forward}$=37$\pm$35, with qualitatively similar $\chi^2_\mathrm{red}$ values ($\mathcal{L}_\mathrm{reverse}$=1.9$\pm$0.76) observed for the other three.

\begin{table}
\centering
\caption{Times and time lag properties of flares observed with \textit{Swift} and TESS}
\begin{tabular}{ccccccccccc}

\hline
\hline
Flare & $\sigma_{NUV}$ & $\sigma_{T}$ & $t_\mathrm{start,~NUV}$ & $t_\mathrm{cntrd,~NUV}$ & $t_\mathrm{stop,~NUV}$ & $t_\mathrm{start,~T}$ & $t_\mathrm{cntrd,~T}$ & $t_\mathrm{stop,~T}$ & $t_\mathrm{lag}$ & $\mathcal{L}_\mathrm{G2/G1}$ \\
~ &  &  & d & d & d & d & d & d & min & \\
\hline
GJ674 F1 & 102 & 4.4 & 60117.07612 & 60117.07773 & 60117.0985 & 60117.07726 & 60117.07707 & 60117.07817 & 0.9 & 0.83 \\
G41-14 F1 & 412 & 28.8 & 59973.09000 & 59973.09187 & 59973.1127 & 59973.09101 & 59973.09066 & 59973.09286 & 1.7 & 0.71 \\
G41-14 F2 & 78 & 6.7 & 59970.29382 & 59970.29548 & 59970.29867 & 59970.29512 & 59970.29457 & 59970.29585 & 1.3 & 0.5 \\
G41-14 F3 & 16.3 & 7.9 & 59973.27434 & 59973.27772 & 59973.28251 & 59973.2774 & 59973.27576 & 59973.2786 & 2.8 & 0.98 \\
G41-14 F4 & 11.6 & 3.6 & 59972.28658 & 59972.28811 & 59972.29027 & 59972.28799 & 59972.28769 & 59972.28822 & 0.6 & 0.78 \\
EVLac F1 & 148 & 9.9 & 59876.50951 & 59876.51106 & 59876.51295 & 59876.51052 & 59876.51028 & 59876.51211 & 1.1 & 0.98 \\
EVLac F2 & 59 & 40.9 & 59877.52189 & 59877.52625 & 59877.53536 & 59877.52574 & 59877.52327 & 59877.5269 & 4.3 & 0.29 \\
EVLac F3 & 34.4 & 4.9 & 59877.64332 & 59877.64429 & 59877.65271 & 59877.64404 & 59877.64397 & 59877.64451 & 0.5 & 1.47 \\
APCol F1 & 119 & 12.3 & 59189.53698 & 59189.54098 & 59189.57371 & 59189.54054 & 59189.54177 & 59189.54454 & 1.1 & -0.12 \\
YZCMi F1 & 111 & 20.4 & 59237.19892 & 59237.20835 & 59237.26318 & 59237.20691 & 59237.20958 & 59237.21825 & 1.8 & 0.34 \\
YZCMi F2 & 37.2 & 6.5 & 59237.26449 & 59237.2664 & 59237.27906 & 59237.26457 & 59237.27097 & 59237.27387 & 6.6 & -0.02 \\
YZCMi F3 & 97 & 11.3 & 59237.27877 & 59237.28008 & 59237.28567 & 59237.28039 & 59237.28063 & 59237.28307 & 0.8 & 2.16 \\
YZCMi F4 & 226 & 23.2 & 59236.40397 & 59236.40541 & 59236.41113 & 59236.40555 & 59236.40654 & 59236.40761 & 1.6 & 1.19 \\
\hline
\end{tabular}
\label{table:flare_timings}
{\newline\newline \textbf{Notes.} Flare timing properties for the NUV and TESS bands. Columns include the flare ID, NUV detection significance, NUV start time, NUV centroid time, NUV end time, TESS start time, TESS centroid time, TESS end time, observed time lag between the NUV and TESS peaks, and the degree to which a two-Gaussian fit to the NUV peak is preferred to a one-Gaussian fit.}
\end{table}

\begin{table}
\centering
\caption{Flare energy budgets of flares observed with \textit{Swift} and TESS}
\begin{tabular}{ccccccc}

\hline
\hline
Flare & $E_\mathrm{NUV}$ & $E_\mathrm{FUV}$ & $E_\mathrm{tot,~T}$ & $E_\mathrm{FWHM,~T}$ & $T_\mathrm{eff}$ & ECF \\
~ & 10$^{30}$ erg & 10$^{30}$ erg & 10$^{30}$ erg & 10$^{30}$ erg & K & \\
\hline
GJ674 F1 & 4.73$\pm$0.11 & 2.2$\pm$1.1 & 2.25$\pm$0.3 & 4.29$\pm$0.53 & 13500 & 4.4 \\
G41-14 F1 & 35.85$\pm$0.86 & 16.4$\pm$8.2 & 36.71$\pm$0.7 & 106.6$\pm$2.2 & 10700 & 2.1 \\
G41-14 F2 & 4.39$\pm$0.11 & 2.0$\pm$1.0 & 3.58$\pm$0.38 & 7.0$\pm$0.84 & 11400 & 2.6 \\
G41-14 F3 & 3.44$\pm$0.91 & 1.6$\pm$0.9 & 6.83$\pm$0.49 & 15.51$\pm$1.06 & 9100 & 1.1 \\
G41-14 F4 & 1.79$\pm$0.43 & 0.82$\pm$0.5 & 0.76$\pm$0.21 & 4.24$\pm$0.61 & 14000 & 4.9 \\
EVLac F1 & 8.36$\pm$0.83 & 3.9$\pm$2.0 & 8.89$\pm$0.39 & 14.18$\pm$1.02 & 10600 & 2.0 \\
EVLac F2 & 15.96$\pm$1.58 & 7.2$\pm$3.7 & 19.62$\pm$0.33 & 165.6$\pm$2.04 & 10200 & 1.7 \\
EVLac F3 & 7.75$\pm$1.32 & 3.8$\pm$1.9 & 1.1$\pm$0.21 & 3.47$\pm$0.6 & 24300 & 14.8 \\
APCol F1 & 38.33$\pm$0.93 & 17.5$\pm$8.8 & 28.4$\pm$0.75 & 88.16$\pm$1.89 & 11700 & 2.8 \\
YZCMi F1 & 63.36$\pm$1.53 & 29.5$\pm$14 & 119.94$\pm$1.14 & 217.89$\pm$2.08 & 9200 & 1.1 \\
YZCMi F2 & 11.63$\pm$1.68 & 5.4$\pm$2.7 & 13.25$\pm$1.01 & 17.84$\pm$1.26 & 10400 & 1.8 \\
YZCMi F3 & 6.85$\pm$0.17 & 3.1$\pm$1.6 & 11.46$\pm$0.56 & 39.66$\pm$1.3 & 9500 & 1.3 \\
YZCMi F4 & 22.59$\pm$0.54 & 10.4$\pm$5.1 & 29.84$\pm$0.49 & 56.31$\pm$0.96 & 10000 & 1.6 \\
\hline
\end{tabular}
\label{table:flare_energies}
{\newline\newline \textbf{Notes.} Properties of the NUV--optical flare energy budgets. Columns include flare ID, total energy measured in the NUV, total energy estimated for the FUV using the \citet{Berger:2024} relations, total energy measured in the TESS band, peak energy measured during the FWHM of the TESS band light curve, $T_\mathrm{eff}$ of the flare, and the empirical correction factor  (i.e. the amount by which the NUV is underestimated by a 9000 K blackbody.)}
\end{table}

\section{Results for the 20 s cadence sample of simultaneous events}\label{simul_flare_results}
Resolving both NUV and optical components at 20 s cadence makes it possible to compare the properties of the flare peaks in each band during the impulsive phase. The impulsive phase is characterized by prompt emission during the rise and peak times of the flare following electron beam heating, and loosely corresponds to the FWHM duration of the TESS flare light curve \citep{Kowalski:2013}. 

\subsection{Illustrations of different NUV-optical energy budgets among flares}\label{indiv_ebudgets}
In the top panels of Figure \ref{fig:twenty_sec_results}, we illustrate two flares in our sample with strong emission at NUV wavelengths but different flux amplitudes in the optical. In the F1 event from G 41-14, the impulsive phase of the TESS counterpart is clearly visible at 20 s cadence. On the other hand, the TESS counterpart of the GJ 674 F1 event is captured at low significance at 20 s cadence and would likely have been missed altogether at 2 min cadence. The difference between these events is due to the fact that G 41-14 is brighter in the NUV but GJ 674 is brighter in the TESS band. However, some flares are intrinsically brighter at NUV wavelengths than others. For example, the F3 event from G 41-14 and F1 event from EV Lac shown in the bottom panels of Figure \ref{fig:twenty_sec_results} have comparable TESS band energies while the NUV energy of the EV Lac flare is 2.4$\times$ higher than that of the G 41-14 flare. More broadly, the NUV:TESS energy ratio of our sample varies from 0.53--7.05, with a median and 1$\sigma$ range of 0.98$\substack{+1.36 \\ -0.22}$.

\subsection{Accurate predictions of the NUV energy from the TESS band}\label{power_law_scalings}
\begin{figure*}
	\centering
        \subfigure
	{
		\includegraphics[trim=10 5 10 0, width=0.48\textwidth]{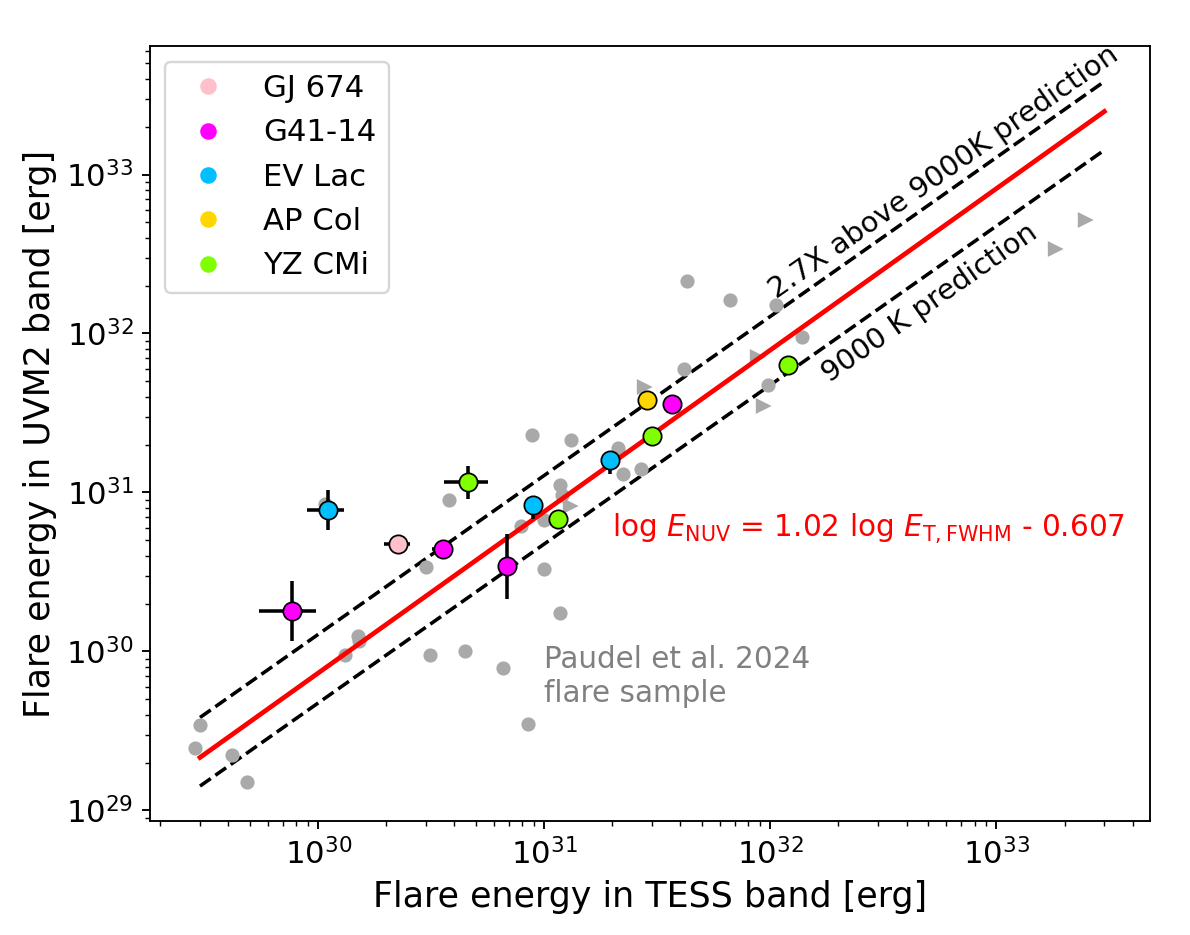}
	}
        \subfigure
	{
		\includegraphics[trim=0 0 0 0, width=0.48\textwidth]{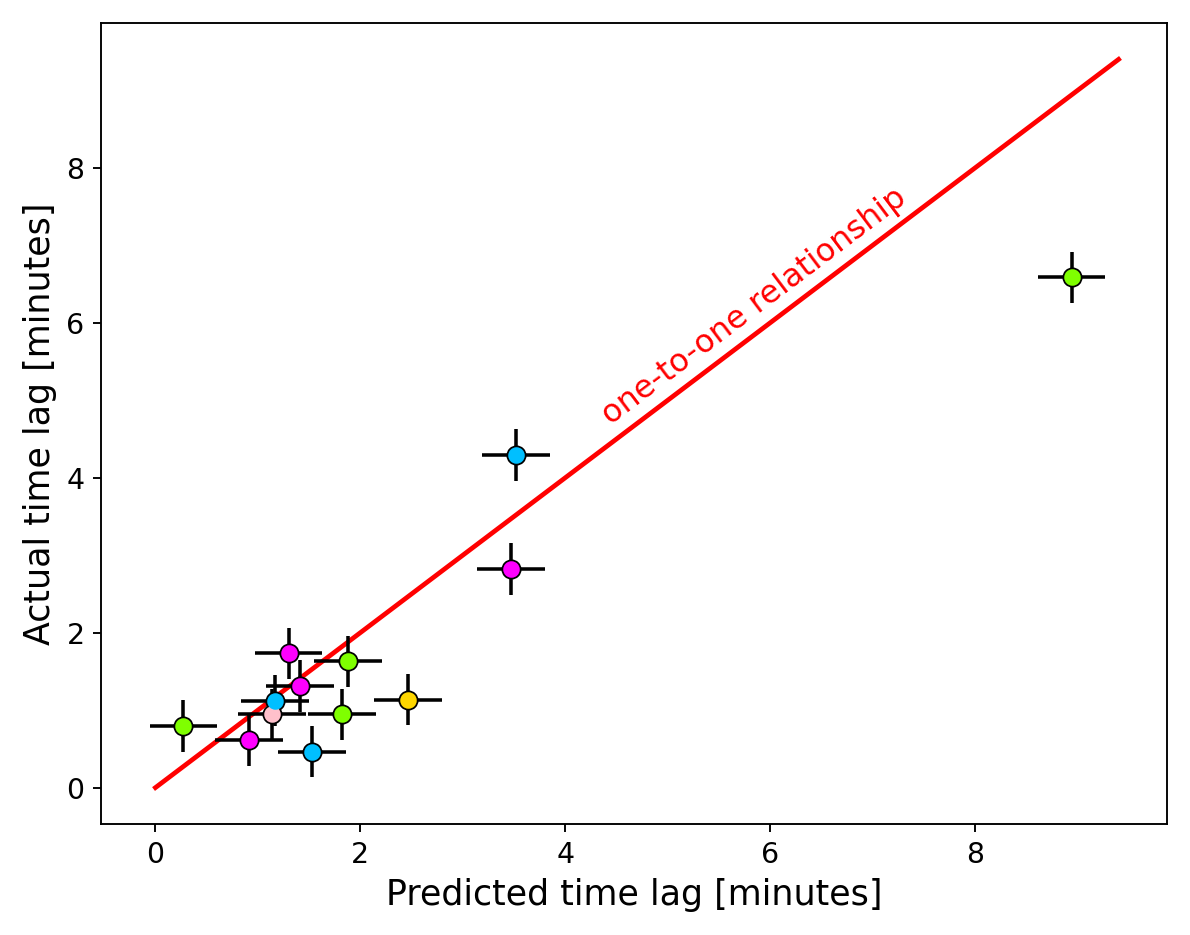}
	}
	\caption{Left: Flare energies in the TESS and NUV bands during the flare peak. Lines representing flare energy budgets for the 9000 K scenario and the NUV-luminous prediction from \citet{Jackman:2022} are shown for reference. The best-fit scaling relationship to the observed sample is shown in red. Right: Predicted time lags between the initial NUV flare peaks and the subsequent TESS band flare peaks. Predicted peak times in the TESS band are determined using the centroid of the second time integral of the flare model fit to the NUV light curves.}
	\label{fig:multiband_results}
\end{figure*}

Differences between the energy budgets of individual flares are one source of the high degree of scatter present in optical--NUV scaling relationships. However, the different timescales and effective temperatures of the hot and cool spectral components during individual flare events also complicate the construction of optical--UV scaling relationships. The entirety of the NUV flare and the FWHM duration of the optical flare are dominated by the hotter component, so we employ the 20 s data to construct an optical--NUV scaling between the FWHM energy of the TESS component and the total energy of the NUV component as shown in Figure \ref{fig:multiband_results}. Minimizing the contribution of the cooler $\sim$5000 K blackbody component from the TESS band energy calculation should therefore provide a better constraint on the total NUV energy of the flare.

To test this, we plot the FWHM energies in the TESS band $E_\mathrm{T,~FWHM}$ against the total NUV flare energies $E_\mathrm{NUV}$ in the left panel of Figure \ref{fig:multiband_results}. The \citet{Paudel:2024} sample is shown for reference in gray and black dashed lines are overlaid that represent the energy budgets of the 9000 K and UV-luminous predictions from \citet{Jackman:2023}. The energy budgets of our larger flares appear to be consistent with those of the \citet{Paudel:2024} sample, while our smaller flares with energies of $E_\mathrm{T,~FWHM}<$5.5$\times$10$^{30}$ erg appear to be UV-luminous. If real, the energy budgets of the small flares may result from optical decoupling that occurs when particle injection traced by prompt UV emission is too weak to produce significant chromospheric heating. Examples of potential optical decoupling have been previously reported during a moderately large flare from Proxima Cen \citep{MacGregor:2021}, the FUV and TESS flare rates of GJ 887 \citep{Loyd:2020}, two flares in the \citet{Brasseur:2023} GALEX and \textit{Kepler} sample, and several events in the \citet{Paudel:2024} sample. However, we consider the energy budgets of the small flares to be tentative due to the low ($\sim$3$\sigma$) signal to noise of the flare peaks. We therefore fit a power law of the form $\log E_\mathrm{NUV}$ = $a_\mathrm{slope} \log E_\mathrm{T,~FWHM} - b$ to the flares with energies $E_\mathrm{T,~FWHM}>$5.5$\times$10$^{30}$ erg and report best-fit coefficients of $a_\mathrm{slope}$=1.02$\pm$0.14 and $b$=0.607$\pm$4.69. The mean scatter or offset of the NUV energies from the values predicted by the best-fit power law for the NUV energies is 28$\pm$7\% for flares of $E_\mathrm{T,~FWHM}>$5.5$\times$10$^{30}$ erg, where the uncertainty in the mean scatter is obtained by recomputing the mean with 1000 bootstrap trials with replacement. For comparison, the mean scatter in the fit residuals for flares of $E_\mathrm{T,~FWHM}>$5.5$\times$10$^{30}$ erg is 56$\pm$11\% for the \citet{Paudel:2024} sample shown in their Figure 14. The factor of 2.0$\pm$0.6 reduction in the mean scatter of the total energy in the NUV may be due to greater similarity between flares during the impulsive phase, but caution is warranted from the small number of flares in our sample.

The best-fit power law for the \citet{Paudel:2024} sample shows a slight preference for a higher ratio of NUV to optical emission at higher energies. We are unable to test this prediction as our sample is primarily made up of events smaller than 1.5$\times$10$^{31}$ erg. In a similar way, the excess UV energies reported in the GALEX and TESS sample of \citet{Jackman:2023} is made up of events of 1.5$\times$10$^{31}$--2.4$\times$10$^{33}$ erg, suggesting we may be missing large UV-luminous flares. Future work in a larger sample of flares is needed to determine if our power law holds at larger energies.

\section{Improvements to the transit detection efficiency for active stars resulting from simultaneous NUV monitoring}\label{transit_detection_efficiency}

\begin{figure*}
	\centering
	{
		\includegraphics[width=0.98\textwidth]{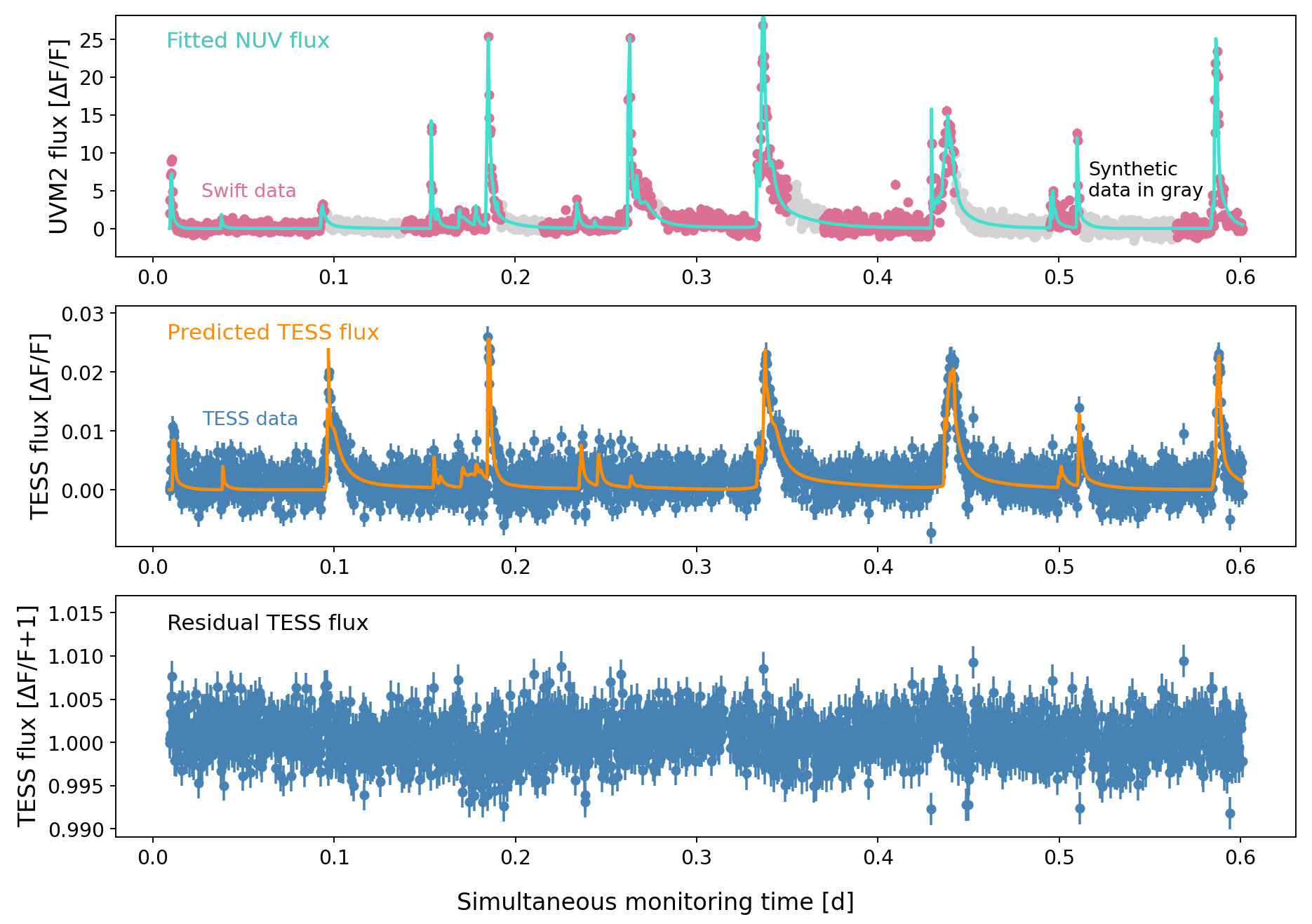}
	}
	
	\caption{Simultaneous 20 s cadence \textit{Swift} NUV and TESS light curve segments have been stitched together to create a time series of 0.6 d. Top panel: The NUV light curve is shown with fluxes and uncertainties adjusted to the photometric uncertainty for the count rate of the faintest star in the sample. Times where the flare decay was not observed are simulated by fitting a flare tail to the observed flux times and shown in gray. The fit to the flares is shown in cyan. Middle panel: The simultaneous data in the stitched TESS light curve is shown with fluxes and uncertainties adjusted to the TESS mag of the faintest star in the sample, AP Col. Small sympathetic peaks during the decay phase of the largest TESS flares are subtracted from the light curve when they occur during the synthetic times in the NUV light curve. This is because they could not have been predicted in principle and would therefore bias the analysis. The model prediction is shown in orange, where the $t_\mathrm{TESS}$ and $\sigma_\mathrm{TESS}$ parameters have been allowed to vary by 30 s and 10\%, respectively. Bottom panel: The residual TESS light curve after subtracting the model of the predicted behavior of the TESS light curve.}
	\label{fig:short_eve_lc_fig}
\end{figure*}

We characterize the improvement seen in transit detection capabilities around young, active stars when simultaneous NUV data are available. We accomplish this by stitching together the individual sets of simultaneous light curve segments to create a longer-duration set of simultaneous light curves. The flux uncertainties of the NUV count rates and TESS fluxes of the segments from the brighter stars in our sample are scaled to those of the faintest star, and the flux values are then randomly varied within the scaled uncertainties. Whenever possible, the light curve segments are stitched together when the count rates and fluxes return to their quiescent values. Several segments include the decay tail of previous flare events that peaked prior to the onset of simultaneous observations. Empirical flare models are fit separately to the decay phase of the earlier events in both bands and subtracted to ensure only the flux from the current flare is included in the stitching process. In a similar way, we use the synthetic NUV data as described above because the tails of the largest flares in our sample continue beyond the NUV observation window. However, these times are covered by the TESS window. The TESS light curves during the times where synthetic NUV data are used sometimes show small peaks superimposed on the large scale shape of the light curve. We therefore fit and remove these small peaks from the TESS band light curve since the synthetic NUV data is not capable of predicting these components. The stitched light curves resulting from this procedure are each 0.6 d in length and contain a range of flare sizes and morphologies. The 0.6 d length is the total observation time obtained from stitching together all available flare observations. The TESS mag of the light curve is equivalent to an active young star of $T$=9.64. The timing and shape of the flares in the stitched TESS band light curve is predicted from the stitched NUV light curve as described in \S \ref{structure_time_lags}. The resulting model parameters of each TESS band flare are then allowed to vary by 10\% to account for small differences between the prediction and actual TESS light curve structure. Finally, the models of the individual flare components are added together to create a model of the entire light curve. The full model is then subtracted from the TESS data to produce a detrended TESS light curve as shown in the bottom panel of Fig. \ref{fig:short_eve_lc_fig}. 

We use Monte Carlo (MC) transit injection and recovery trials to determine the degree of improvement in detection efficiency between the original and detrended TESS light curve. The short duration of the 0.6 d light curve prevents the injection of multiple transits for most realistic orbital periods and transit durations. We therefore divide the 0.6 d light curve up into 10 sections based on times where the flare model has returned to quiescence. We create 1000 unique light curves of 30 d each with 510 sequential random draws and assemble the sections according to the draw order. In each of 10,000 MC trials, one of these 30 d light curves is randomly selected and a transiting planet is injected. Each transit model is computed using the \texttt{batman} code, where the orbital period and radius of the planet are drawn from uniform distributions in orbital period and transit depth parameter space, with orbital periods of 1 to 15 d and transit depths of 0.5 to 5$R_\oplus$. We compute the semi-major axis of each planet from Kepler's third law for the mass of a K7V host star, assume zero eccentricity, and assume quadratic limb darkening coefficients of 0.17 and 0.26. The resulting transit model is injected separately into the original and detrended versions of the light curve and Box Least Squares (BLS) periodograms are computed over a range of periods from $\sim$0.5--15 d. We use the BLS \texttt{autopower} method in \texttt{astropy} and test transit durations from 0.05--0.2 d with a frequency factor to 10. A linear trend is fit to each BLS periodogram and subtracted to remove the increasing noise floor systematic. We also compute the standard deviation (STD) in power for periods within a 20\% region around each injected period in order to estimate the significance of the detection and to normalize the periodograms of the original and detrended light curves onto the same y-axis. The 20\% region was selected by visual inspection and chosen to ensure the STD calculation contains typical noise spikes for injected periods from 1--15 d.

\begin{figure*}
	\centering
	{
		\includegraphics[width=0.98\textwidth]{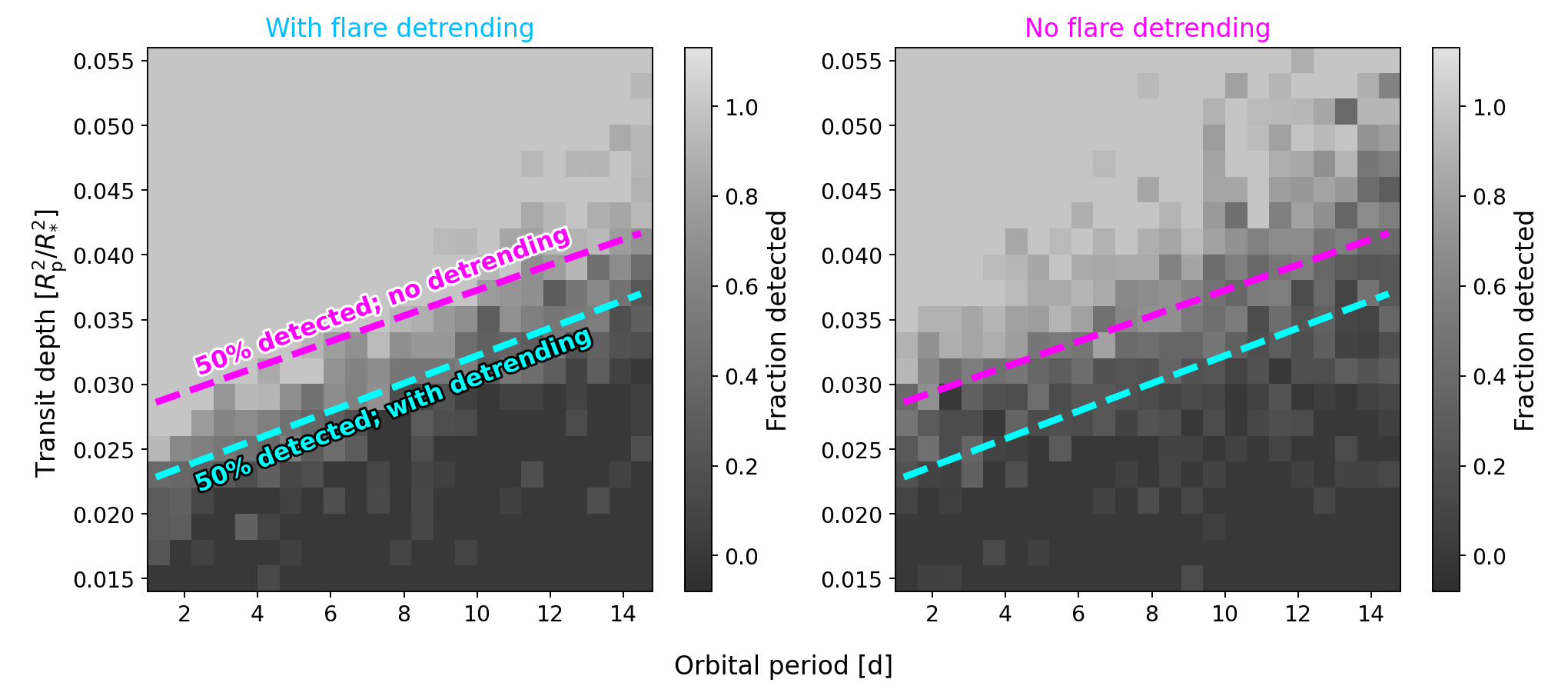}
	}
	
	\caption{Left: Transit detection efficiency for a $T$=9.64 star after flare detrending computed from 10,000 planet injection and recovery trials. The grayscale grid is the fraction of transits recovered in each planet period and transit depth bin, with lighter shades representing higher detection fractions. The 50\% detection threshold with detrending is highlighted in cyan, while the 50\% detection threshold in the case of no flare detrending is shown in magenta for comparison. Over-plotted are simulated planets with a colormap representing their SNR in the BLS periodogram. Right: Same as the left panel for the no flare detrending scenario. The flare detrending process increases the transit depth $\delta$ by $\Delta\delta$=0.0052, which corresponds to an increase of 0.42$R_\oplus$ for a K7V star.}
	\label{fig:transit_efficiency}
\end{figure*}

Two-dimensional histograms of the fraction of 10,000 injected planets with 5$\sigma$ detections are computed for both the original and detrended scenarios to assess the improvement in planet yields due to flare detrending. These histograms are shown in grayscale in Figure \ref{fig:transit_efficiency}, where lighter shades represent higher probabilities of detection. The SNR of the peak in the BLS periodogram is shown for a randomly selected subset of planets, with yellow representing higher SNR. We over-plot lines representing the 50\% detection threshold in period and transit depth space to facilitate the comparison. The detection probability varies as a function of both the transit depth and orbital period, and is highest for large planets and at short periods.

\section{Multiwavelength flare yields with the Early Evolution Explorer}\label{applic_eve}
We explore the instrument requirements for a NASA Astrophysics Small Explorer concept capable of rejecting the null hypothesis that the NUV energies of flares from young stars are consistent with a 9000 K blackbody.

\begin{figure*}
	\centering
	{
		\includegraphics[width=0.99\textwidth]{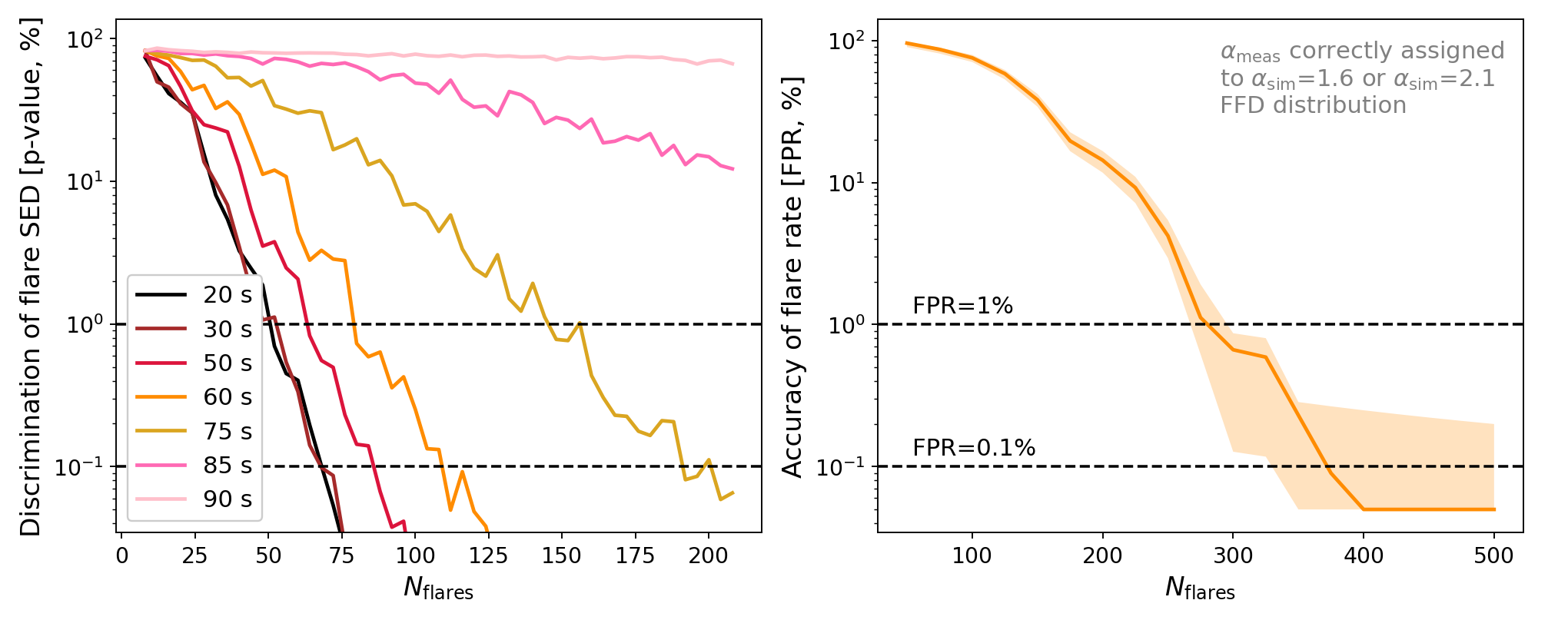}
	}
	
	\caption{Left panel: The number of flares needed to reject the $\sim$9000 K energy budget null hypothesis at a given confidence level using an A-D test. Curves are shown for a range of cadences, where each curve represents a 3$\sigma$ confidence interval on the distribution of p-values measured from a sample size of $N_\mathrm{flares}$. Right panel: The number of flares needed to ensure the observed FFD power law slope slope $\alpha_\mathrm{FFD}$ is representative of the true average flare rate in both the NUV and optical bands. This is the number of flares needed to reject the null hypothesis of $\alpha_\mathrm{FFD}\approx$1.6, and is insensitive to cadences of 20--120 s.}
	\label{fig:eve_required_yields}
\end{figure*}

\subsection{Number of flares needed to test optical-UV scaling relations}\label{sim_yield_reqs}
The first step is to determine how many simultaneous flare detections are required to reject the 9000 K hypothesis at the 99\% and 99.9\% confidence levels. We perform 2000 draws of two flare samples composed of $N_\mathrm{flares}$ each, in which the two samples represent flare energy budgets consistent with a 9000 K blackbody and the UV-luminous predictions from \citet{Jackman:2023}. The TESS energies are identical between the two distributions and are sampled from a simulated flare frequency distribution (FFD) of $\alpha_\mathrm{FFD}\approx$2. The NUV energies of the flares are obtained by scaling the TESS energies by factors of 0.69$\pm$0.6 and 1.86$\pm$0.6 for the 9000 K and UV-luminous predictions, respectively. The two-sample Anderson-Darling (A-D) test statistic and p-value are computed for each draw and the 3$\sigma$ upper limit on the range of p-values is recorded and shown in the left panel of Figure \ref{fig:eve_required_yields}. We find 50 and 70 flares are needed to reject the 9000 K hypothesis at the 99\% and 99.9\% level, respectively.

In addition to the ``clean" two-sample test described above, we also compute the same A-D tests for contaminated samples. Long cadences bias the $F_\mathrm{NUV}/F_\mathrm{TESS}$ ratio of 9000 K flares toward higher values given the short duration of flare peaks at NUV wavelengths. We therefore simulate simultaneous NUV and TESS band light curves of 2000 flares at 1 s cadence using the \citet{Tovar_Mendoza:2022} template and set $F_{NUV}/F_{TESS}$=0.475 for each event. We bin the flares down to cadences of 5, 10, 20, 30, 40, 60, 90, and 120 s. Similarly, we simulate 2000 flares at 1 s cadence and set $F_\mathrm{NUV}/F_\mathrm{TESS}$=1.283 before binning the flares down to the same set of cadences. Even though some flares are UV-luminous at the $F_\mathrm{NUV}/F_\mathrm{TESS}>$6.6 level, we set the value for the hypothesis test at 1.283 to ensure we can distinguish UV-luminous events under the more difficult conditions allowed by the current literature range of values. In each simulated flare, the FWHM of the NUV flare is drawn from a Gaussian distribution with $\mu_\mathrm{FWHM}$=90 and $\sigma_\mathrm{FWHM}$=30 s. The FWHM of the TESS flare is obtained from the time integral of the NUV light curve following the procedure described in \S \ref{structure_time_lags}. The FWHM duration of the binned light curves is measured in each band for each cadence and the energy ratio within the FWHM at the tested cadence is recorded. The fraction of trials in which the observed ratio cannot be distinguished at the 3$\sigma$ level between the $F_\mathrm{NUV}/F_\mathrm{TESS}$=0.475 and $F_\mathrm{NUV}/F_\mathrm{TESS}$=1.283 distributions increases as the observing cadence approaches the FWHM duration. We construct a FPR-versus-cadence relation, where the FPR is the fraction of trials in which a 9000 K flare would be mistaken for a UV-luminous flare at a given cadence. We then add contamination to each clean sample of $N_\mathrm{flares}$ using our FPR-vs-cadence relation by swapping $N_\mathrm{flares}\times$FPR randomly selected flares between the 9000 K and UV-luminous samples. The A-D test statistic is recomputed on the contaminated samples for FPR fractions of 0.3\%, 0.6\%, 1\%, 5\%, 10\%, 20\%, 30\%, 40\%, and 50\%. These values are then mapped to the respective cadences of 20, 30, 35, 50, 60, 75, 85, 90, and 120 s. While a 0.3\% FPR corresponding to 20 s cadence requires the same number of flares as for the clean sample, a 40\% FPR corresponding to 90 s cadence is unable to reject the null hypothesis at the 99\% level for even $N_\mathrm{flares}<$1250 events.

Next, we determine how many flares are needed to ensure the slope of the cumulative FFD is representative of the true power law slope of the underlying population. The cumulative FFD is a power-law relationship of the form $\log{\nu}=(1-\alpha)\log{E} + \beta$ that describes the rate at which flares of a given energy or larger occur per unit time. Here, $\nu$ is the flare rate d$^{-1}$ for events with an energy greater than or equal to $E$ erg, 1-$\alpha$ is the power law index that determines the relative distribution of small versus large flares, and $\beta$ describes the overall flare activity level. While $\alpha$ remains relatively constant with stellar mass, $\beta$ increases with mass due to the increased depth of the convection layer. The power law slope $\alpha_\mathrm{FFD}$ determines whether frequent small flares or rare superflares dominate the contribution of flares to the radiation environments of orbiting planets. 

It is unknown if shallow power law indices observed for young stars in large optical surveys (e.g. \citealt{Feinstein:2020, Feinstein:2022a, Feinstein:2024}) correspond to shallow slopes at NUV wavelengths \citep{Jackman:2023, Paudel:2024}. We determine the minimum number of flares needed to distinguish between shallow ($\alpha_\mathrm{slope}$=1.6; \citet{Feinstein:2020}) and steep ($\alpha_\mathrm{slope}$=2.1, selected to lie across the $\alpha$=2 boundary) FFD slopes at 99\% and 99.9\% confidence levels with Monte Carlo simulations of simulated versus observed $\alpha_\mathrm{FFD}$ values as a function of $N_\mathrm{flares}$. We draw 2000 samples of $N_\mathrm{flares}$ each from simulated FFDs with power law indices of $\alpha_\mathrm{FFD}$=1.6 and 2.1 for the shallow and steep scenarios, respectively. The choice of $\alpha_\mathrm{FFD}$=1.6 is set by the shallow values measured from young star flares in TESS by \citet{Feinstein:2020}, while the choice of $\alpha_\mathrm{FFD}$=2.1 ensures the largest flares dominate the flare radiation environment in the band. For each sample of $N_\mathrm{flares}$, the cumulative FFD is measured separately for the flares drawn from the shallow and steep distributions and the two $\alpha_\mathrm{FFD}$ distributions are compared. A Gaussian function is fit to each of the two $\alpha_\mathrm{FFD}$ distributions of slopes to measure the mean and standard deviation. We then measure the fraction of $\alpha_\mathrm{FFD}$ values drawn from one distribution that lie within 3 standard deviations of the other distribution's mean, and vice versa. The higher of the two FPR values for each $N_\mathrm{flares}$ value is recorded for all $N_\mathrm{flares}$ tested as shown in the right panel of Figure \ref{fig:eve_required_yields}. We find that 280 flares are needed for a 99\% confidence interval (CI) and 370 flares are needed for a 99.9\% CI. Contamination is also introduced as for the energy budget test, but no difference is observed in the number of flares required for cadences of 120 s or less.

\subsection{Estimating flare rates in the dense core regions of young clusters}\label{core_flare_yield}
\begin{figure*}
	\centering
	{
		\includegraphics[width=0.91\textwidth]{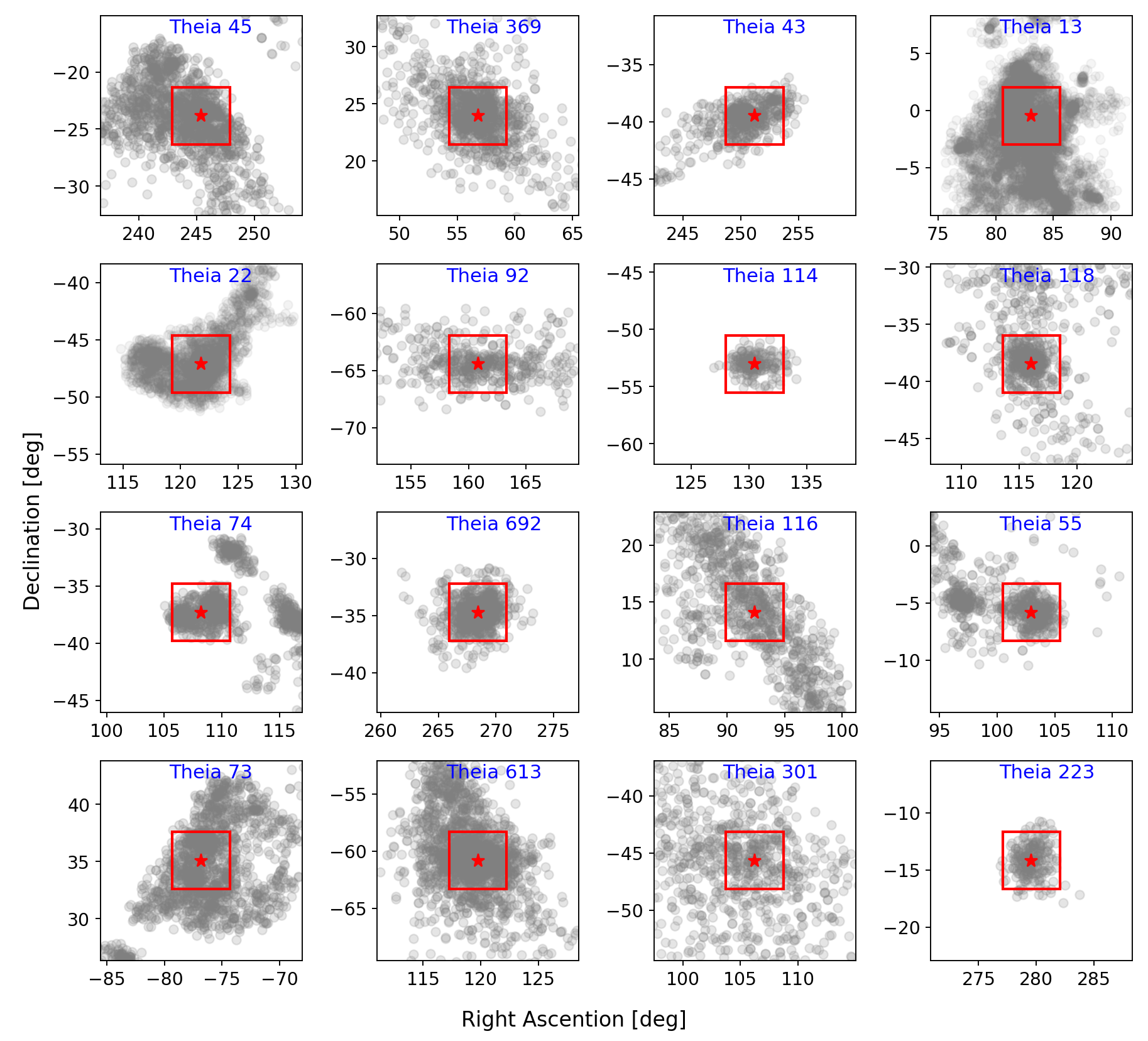}
	}
	
	\caption{Examples of 5$\times$5 deg$^2$ cluster pointings that produce high yields of simultaneous NUV flare detections. Here, we show the core region with the highest flare yield per cluster even though some clusters have multiple high-yield core regions.}
	\label{fig:eve_core_pointings}
\end{figure*}

The second step is to determine the flare rates in the dense core regions of nearby young clusters. Relatively few TESS programs to date have obtained comprehensive observations of young clusters at ages of 5--100 Myr with 20 s cadence observations in order to explore the variation in the flare rates and properties with stellar age. A cluster's flare rate depends on the number density of low mass stars in the core region, stellar age, and membership contamination fraction. We begin by collating a list of the densest core regions in nearby clusters from the \textit{Theia} catalog \citep{Kounkel_Covey:2019} and then supplement the membership lists with other catalogs available in the literature to produce an inclusive list (e.g. \citealt{Kounkel_Covey:2019, Kounkel:2020, Ujjwal:2020, Moranta:2022}). The updated catalog is then down-selected to clusters within $\sim$400 pc, ages of 5--400 Myr, and that contain at least 200 candidate members within a 5 deg radius of its densest region. These criteria exclude the majority of clusters that produce low flare yields. The top regions of enhanced density are identified by visual inspection of each remaining cluster and checked against the number of K7--K9 and M0--M6 dwarfs in a nominal 5$\times$5 deg$^2$ box around the coordinates of the densest region. When a cluster contains multiple core regions, the top three performing regions are included as candidate pointings based on their yield estimates. In total, we identify 114 candidate pointings within 76 clusters that meet our criteria, including a number of well-known groupings such as Orion, Centaurus A, Upper Scorpius, AB Dor, IC 2602, and the Vela OB2 region. We illustrate candidate pointings for the highest-yield core regions in the top 16 clusters in Figure \ref{fig:eve_core_pointings} and list the properties of all 114 regions and their respective flare yields in Table \ref{table:cluster_pointings}.

The core region average flare rate is measured by aggregating all sources with 2 min cadence TESS light curves with ages of 15--60 Myr across our 76 core regions. Flare rates of low-mass stars remain fairly stable for $\sim$20--200 Myr before decreasing with age \citep{Feinstein:2020, Ilin:2021, Feinstein:2024}. We therefore set conservative upper and lower age limits of 15 and 60 Myr to avoid biasing the flare rates with less active sources. In total, TESS observed 21 K5--K9 dwarfs, 26 M0--M2 dwarfs, and 6 M3--M5 dwarfs between Sectors 1--71. We download \texttt{SAP\_FLUX} short cadence light curves for each of these sources from MAST, and verify the GO programs responsible for requesting these stars did not select targets based on prior knowledge of their stellar activity levels. We use average flare rates measured within the core regions we intend to observe with EVE rather than average rates measured across all kinematically-selected members of similar age across the sky. This is because field stars with similar kinematic properties in the peripheral regions of some associations are more difficult to distinguish from bona fide members \citep{Kounkel_Covey:2019, Bouma:2021}, resulting in a higher false positive rate (FPR) and hence lower average flare rate.

\begin{table}
\centering
\caption{Flare yields for candidate young cluster pointings}
\begin{tabular}{cccccccc}
\hline
\hline
Pointing & R.A. & Dec & Age & $N_\mathrm{dK}$ & $N_\mathrm{dM}$ & $n_\mathrm{fl,~TESS}$ & $n_\mathrm{fl,~NUV}$ \\
 & ddeg & ddeg & Myr &  &  &  & \\
\hline
Theia 45-1 & 245.390 & -23.830 & 15 & 137 & 617 & 1412.2 & 104.9 \\ 
Theia 369-1 & 56.800 & 23.960 & 130 & 72 & 670 & 908.2 & 88.3 \\ 
Theia 45-2 & 240.370 & -23.800 & 15 & 63 & 226 & 752.6 & 56.5 \\ 
Theia 43-1 & 251.200 & -39.470 & 16 & 82 & 440 & 617.5 & 44.7 \\ 
Theia 692-1 & 268.429 & -34.710 & 390 & 319 & 1148 & 570.9 & 41.8 \\ 
Theia 43-2 & 191.650 & -56.440 & 16 & 27 & 73 & 275.5 & 39.0 \\ 
Theia 45-3 & 242.180 & -18.800 & 15 & 47 & 167 & 601.8 & 37.6 \\ 
Theia 13-1 & 83.095 & -0.470 & 7 & 1077 & 3308 & 1184.5 & 35.8 \\ 
Theia 13-2 & 84.380 & -6.130 & 7 & 1050 & 1178 & 885.5 & 30.0 \\ 
Theia 22-1 & 121.790 & -47.130 & 12 & 402 & 1360 & 752.2 & 29.1 \\ 
Theia 613-1 & 119.750 & -60.830 & 250 & 424 & 1894 & 508.6 & 27.8 \\ 
Theia 92-1 & 160.820 & -64.440 & 35 & 30 & 225 & 254.6 & 23.1 \\ 
Theia 43-3 & 283.350 & -36.590 & 16 & 21 & 118 & 298.3 & 20.1 \\ 
 &  &  &  & ... &  &  & \\
\hline
\end{tabular}
\label{table:cluster_pointings}
{\newline\newline \textbf{Notes.} Parameters of 114 candidate pointings of 25 deg$^2$ across 76 clusters, given in descending order of NUV flare yield. Only a portion of the table is shown here, which is available in its entirety in machine-readable form. Columns are pointing ID, field center coordinates, age, number of K dwarfs, number of M dwarfs, TESS-band flare yield per 30 d stare time, and NUV flare yield per 30 d stare time. Pointings are named by Theia catalog number and subfield designation.}
\end{table}

Flares are identified in the 2 min cadence light curves using the same flare-finding algorithm and visual confirmation process described in \S \ref{sec:tess_obs}. The energy of each flare in the TESS bandpass is also measured as described above and the total observation time is recorded for each target. We combine the flares and target monitoring times together by stellar classification in order to produce averaged flare rates for K5--M0, M0--M2, and M2--M4 stars following the procedure described in \citet{Howard:2019}. We compute the cumulative FFD for an ``average" star in each of our three spectral class bins. We report FFDs of $\log{\nu_\mathrm{TESS}}$ = -0.84 $\log{E_\mathrm{TESS}}$ + 27.34 for the early M dwarfs and $\log{\nu_\mathrm{TESS}}$ = -0.85 $\log{E_\mathrm{TESS}}$ + 26.81 for the mid dwarfs. The late K dwarf FFD is comparable to that of the early M dwarfs so we use the early M dwarf FFD for both groups since M dwarfs dominate the flare yield calculation. Finally, we estimate the NUV flare rate for each stellar mass and age combination by scaling the TESS energies into the NUV band using a 9000 K blackbody since the NUV flare rate will be lower for the 9000 K scenario than for the UV-luminous scenario.

The occurrence of flares with a fractional flux amplitude of $A_\mathrm{TESS}$ or greater at 20 s cadence is estimated from each FFD in order to identify the smallest detectable flare for a given star as a function of its TESS mag. To achieve this, we fit a power-law relationship in log-log space to the TESS band energies and flux amplitudes of 2575 flares from mid M dwarfs reported in 20 s TESS observations by \citet{Howard_MacGregor:2022}. This approach provides a best-fit scaling of log $A_\mathrm{TESS}$ = 0.69 log $E_\mathrm{TESS}$ - 24.41 for the 20 s cadence mid M sample. The stellar contrast of a flare of $E_\mathrm{TESS}$ decreases with increasing stellar brightness, requiring separate conversions for late K dwarfs, early M dwarfs, and mid M-dwarfs. We scale our mid M energy-to-amplitude relationship by the relative bolometric luminosities of M0--M1 and K7--K9 dwarfs given in \citet{Kraus_Hillenbrand:2007} to obtain reduction factors for the flare amplitudes of 0.333 and 0.132, respectively. We also verified the accuracy of the early M dwarf scaling by direct comparison with the energies and flux amplitude distribution for 539 flares from the early M dwarfs.

\subsection{The impact of instrument trades on mission flare yields}\label{instrument_trades}
The final step is to determine the flare yields expected for a range of mission concepts that we consider to be feasible given the mass and cost limits of the NASA Astrophysics Small Explorers program. We therefore carry out a trade study for concepts with a 2--3 year primary mission, low earth orbit, and apertures of 20--34 cm. The choice of aperture, detector sensitivity, and field of view are the primary drivers for the flare yield. We determine the number of flares expected over the mission lifetime by varying each of these parameters. Flare yields are calculated for a nominal mission lifetime of two years with half of that time reserved for science observations and the other half for operations and science margin.

\subsubsection{Trade study assumptions}\label{trade_assumptions}
Our trade study assumes an overall throughput of 9\% in the NUV band and 29\% in the red optical band. We note these are conservative assumptions and these values may be different than those of the final optical and detector design. Our trade study includes effective apertures from 20--34 cm in increments of 2 cm, dark currents of 0.05, 0.1, 0.2, and 0.4 $e^{-}$ s$^{-1}$, and read noise values of 0.5, 1, 2, 4 e$^{-}$. Radiometry SNR versus brightness curves for our trade study are computed for the above optical throughput assuming noise sources including stellar shot noise, dark current, read noise, and sky background. Since flares are thought to exhibit $T_\mathrm{eff}$=9000 K, our SNR versus brightness curves in each band assume the spectrum of HD 104237, an A4V star with a TESS mag of 6.346$\pm$0.006 and distance of 106.6$\pm$0.5 pc. In the NUV band, the stellar signal is 956,000 $e^-$ and the shot noise is 980 $e^{-}$. In the red optical band, the stellar signal is 24,700,000 $e^-$ and the shot noise is 4,970 $e^{-}$. At NUV wavelengths, $\sim$97\% of the sky foreground is due to geocoronal emission from O+ ions, producing 2.2 $e^{-}$ s$^{-1}$ pix$^{-1}$ at an altitude of 580 km and for $\sim$10" pixels. We scale the overall SNR of the A4V source with stellar distance to obtain the SNR versus brightness curves for both bands as a function of TESS magnitude.

\subsubsection{Trade study yield calculations}\label{trade_calcs}
\begin{figure*}
	\centering
        \subfigure
	{
		\includegraphics[trim=40 0 10 20, width=0.5\textwidth]{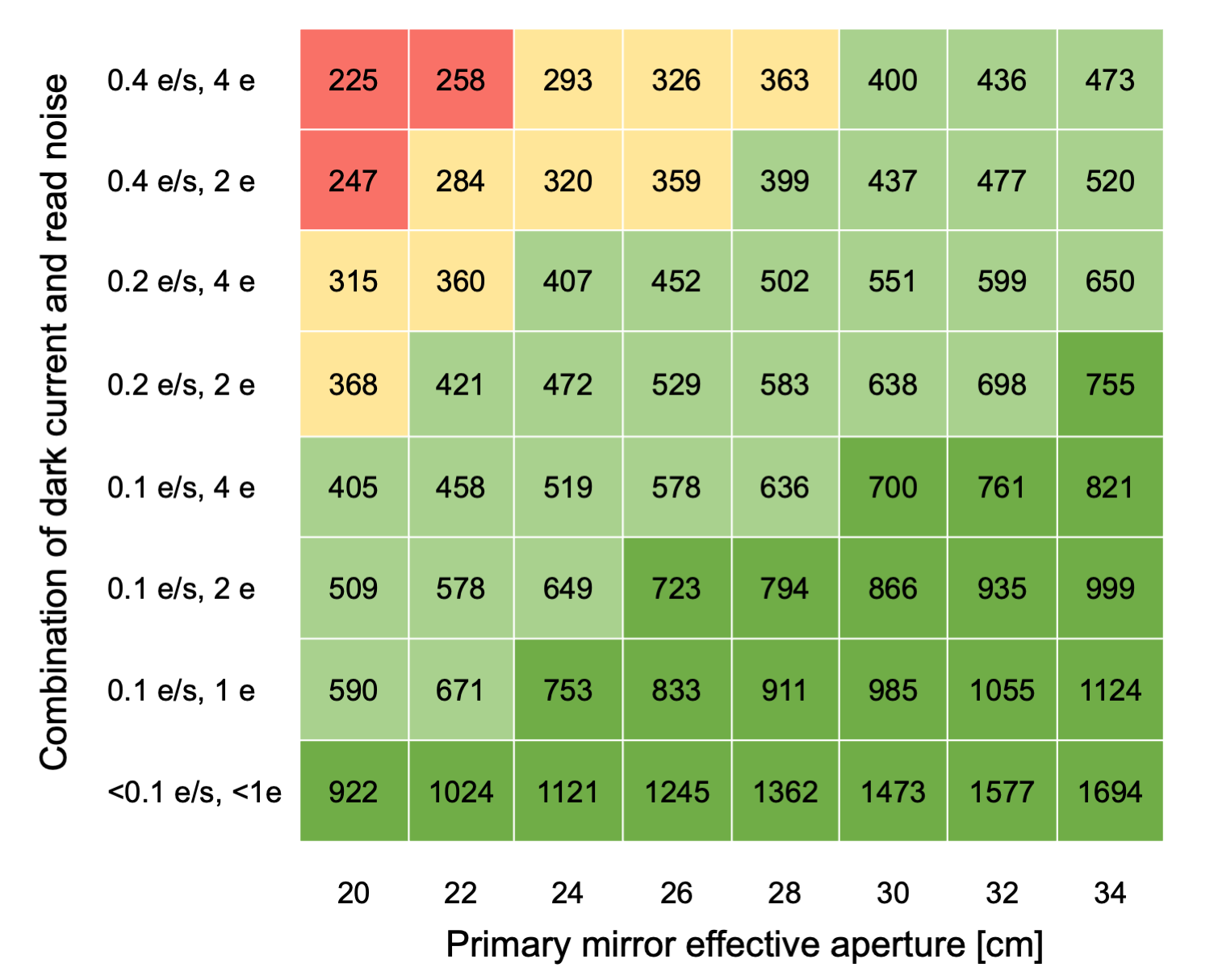}
	}
        \subfigure
	{
		\includegraphics[trim=5 0 35 0, width=0.45\textwidth]{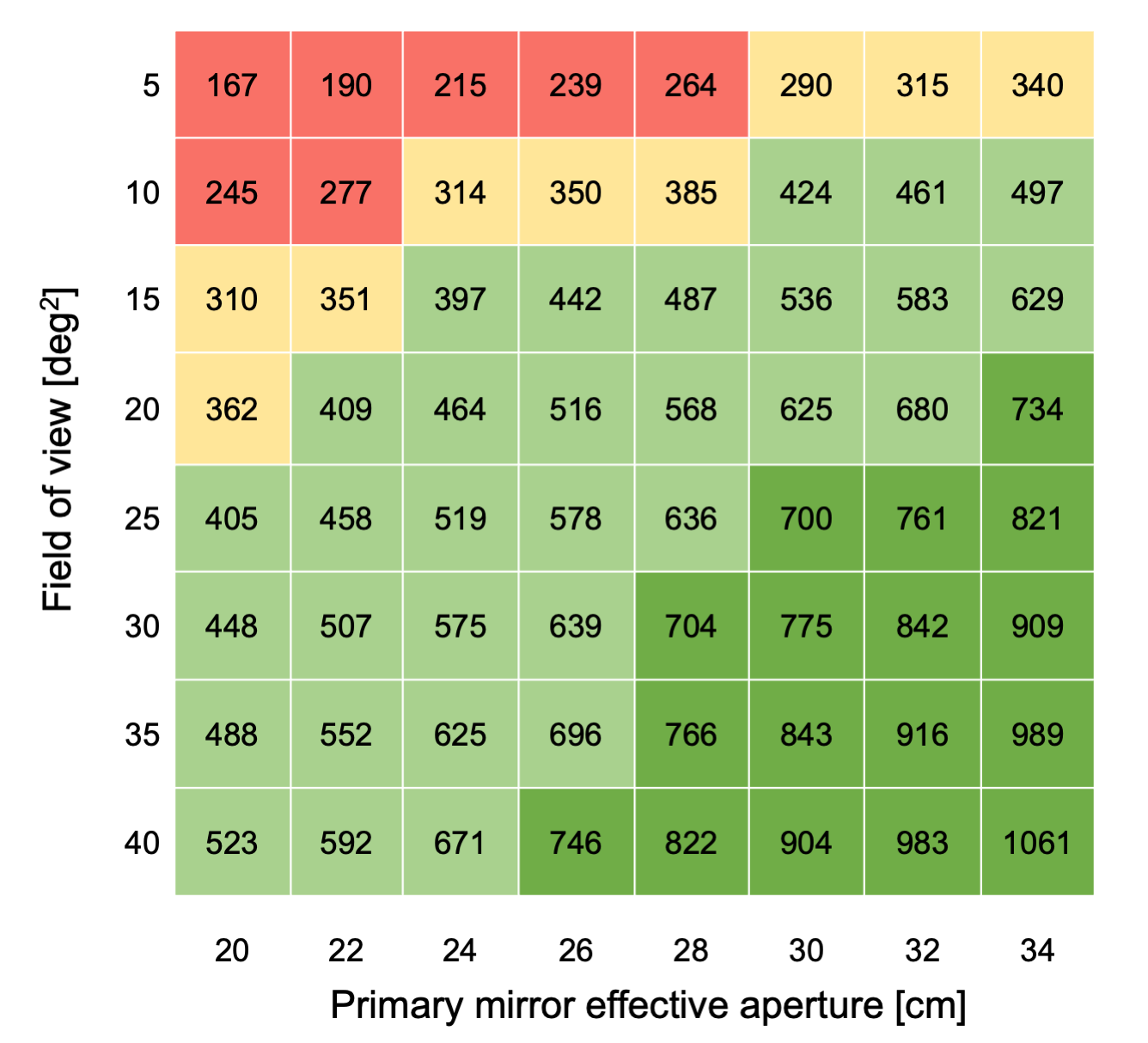}
	}
	\caption{Multi-wavelength flare yield estimates for a two year science mission, with conservative assumptions of 50\% science margin and 9\% NUV throughput. The FoV plot assumes 0.1$e^{-}$ s$^{-1}$ and a read noise 4$e^{-1}$. Flares are detected at 3$\sigma$. This figure is intended to illustrate how large differences in flare yields can arise when varying the instrument parameters across a reasonable range of values, and is not intended to specify the final instrument design of EVE. Other sets of instrument parameters are possible (e.g. increased transmission, quantum efficiency, or aperture), but this study demonstrates acceptable flare yields are achieved for feasible instrument designs.}
	\label{fig:trade_study_yields}
\end{figure*}

The flare yield is computed for each of the 114 candidate pointings identified in \S \ref{core_flare_yield} using the SNR versus TESS mag radiometry curves for each band, the number of late K, early M, and mid M stars in the FoV, and the flux-amplitude flare rates. A stare time of 30 d is assumed for each pointing. The number of flares exceeding 3$\sigma$ above the local photometric noise in each band is computed on a star-by-star basis since the rate of detectable events is a function of the TESS mag and stellar mass. The combined yield for each pointing $N_\mathrm{pointing}$ is obtained for each band by summing over the yields of the individual stars in the FoV according to Eq. \ref{eq:fl_yields}:
\begin{equation}
\label{eq:fl_yields}
    N_\mathrm{pointing} = \sum_{n=1}^{n_\mathrm{stars}(FoV)} \nu_{\mathrm{ampl},{3\sigma}}(m_\mathrm{n}, T_\mathrm{n}) \times t_\mathrm{obs}
\end{equation}
Here, $\nu_{\mathrm{ampl},{3\sigma}}$ is the rate at which flares occur with an amplitude exceeding a 3$\sigma$ detection for a given SNR curve, $m_\mathrm{n}$ is the stellar mass, and $T_\mathrm{n}$ is the TESS mag. The yields obtained for all 114 separate pointings are then sorted in descending order and the top 12 pointings in clusters with ages less than 100 Myr are added together to produce a total flare yield over the mission lifetime, $N_\mathrm{flares}$ = $\sum_{p}^{114} N_\mathrm{pointing}(p)$. We compute a grid of $N_\mathrm{flares}$ predictions for each combination of effective aperture, dark current, and read noise shown in the left panel of Fig. \ref{fig:trade_study_yields}. Here, the FoV is fixed to 25 deg$^2$. We also compute a grid of $N_\mathrm{flares}$ predictions for each aperture and FoV shown in the right panel of Fig. \ref{fig:trade_study_yields}. Here, the dark current and read noise are set to 0.1 $e^-$ s$^{-1}$ and 4 $e^-$, respectively.

\subsubsection{The impact of extinction on the NUV flare yield}\label{nuv_extinct}
The 12 top-ranked fields below 100 Myr are each in clusters with mean distances of 150--200 pc and estimated $A_\mathrm{V}\sim$0.1 (i.e. Theia 43, 45, 92, 114, 118), with the exception of two fields in Theia 13 (Orion) and one in Theia 22 (Vela OB2). The S/N of NUV flares is computed from the observed TESS magnitude and a 9000 K blackbody flare spectrum. Scaling from the TESS band into the NUV in this way implicitly assumes the extinction in the observed TESS magnitude also holds in the NUV band. This assumption gives negligible errors in the NUV for $A_\mathrm{V}\sim$0.1 and $R_\mathrm{V}$=3.1, typical of many regions in Sco Cen and other nearby groups (e.g. \citealt{Luhman:2022, Luhman:2023, Ratzenbock:2023}). However, the \citet{Schlegel:1998} extinction maps indicate this assumption does overestimate the S/N at NUV wavelengths by factors of up to 1.4 ($A_\mathrm{V}$=0.45) and 1.6 ($A_\mathrm{V}$=0.61) for the Theia 13 and 22 fields, respectively. We consider these factors to be upper limits because the extinction maps are integrated along the line of sight and thus include extinction at greater distances than the TIC sources, which we verify for Theia 13 and 22 by comparing the distribution of observed TESS magnitudes for stars with coordinates overlapping high extinction ($A_\mathrm{V}>$2) regions and nearby stars that do not overlap the extinction enhancements. The comparisons show the mean TESS mag of stars overlapping high regions only decreases by $\sim$0.3 mag, consistent with $A_\mathrm{V}$=0.5. Decreasing the S/N by these upper-limit factors of 1.4 and 1.6 reduces the NUV yields by 40\% and 50\% for the Theia 13 and 22 fields, respectively. Summed across all 12 top-ranked fields and assuming a stare time of 30 d per field, this drop corresponds to a decrease of 8.7\%, from 460 to 420 NUV flares for the 25 deg$^2$, 0.1 $e^{-}$ s$^{-1}$, and 4$e^-$ scenario. As a result, we do not include the 8.7\% drop in the reported yields arising from the upper limits on extinction.

In Figure \ref{fig:trade_study_yields}, flare yields that are not capable of distinguishing between the 9000 K and UV-luminous scenarios at the 3 sigma level and the steep versus shallow slopes at the 1\% level are shaded in red. Most combinations of optical and detector properties enable the mission to reject the 9000 K null hypothesis at $\geq$3$\sigma$ and to reject the shallow power law slope for the flare rate hypothesis in both bands at a p-value of $\geq$0.01. Flare yields that are capable of testing these hypotheses when binning over all stellar ages and masses are shown in yellow. Flare yields that are capable of testing these hypotheses at the 0.1\% level when binning over all stellar ages and masses are shown in light green. Flare yields that are capable of testing these hypotheses at the 0.1\% level without binning over all stellar ages and masses are shown in dark green. Only the 20--22 cm effective aperture and 0.4$e^-$ s$^{-1}$ dark current combinations are unable to reject the null hypotheses at these levels in 12 months of science time given the conservative throughput values we assume. Similarly, only apertures of 20--28 cm and FoV values below 10 deg$^2$ fail to reject the null hypotheses at these levels in 12 months of science time.

\section{Discussion and Conclusion}\label{discuss_conclude}
We present new 20 s cadence simultaneous observations of flares at NUV and optical wavelengths from GJ 674, G 41-14, and EV Lac obtained with \textit{Swift} and TESS in order to characterize the time lag and energy budget of the flare peak. We extend our sample with new analyses of other 20 s cadence \textit{Swift} and TESS flares from EV Lac \citep{Inoue:2024}, AP Col, and YZ CMi \citep{Paudel:2024} for a total sample of 13 events. The 20 s cadence data enables us to characterize the time lag between the NUV and TESS band peaks for the first time, demonstrating the TESS peak of most flares occurs after the NUV peak. The time lag is less than 2 min for 85\% of the sample, but lags from 0.46--6.6 min are observed. We further demonstrate the time lag is correlated with the shape of the flare light curves and can be empirically predicted to a 36$\pm$30\% accuracy using the second integral with respect to time of the NUV light curve. Furthermore, the shape of the flare light curve in the NUV and TESS bands appears to follow a relationship reminiscent of the Neupert effect observed between the UV and SXR flare light curves.

\begin{figure}
  \begin{minipage}[c]{0.59\textwidth}
    \includegraphics[scale=0.59]{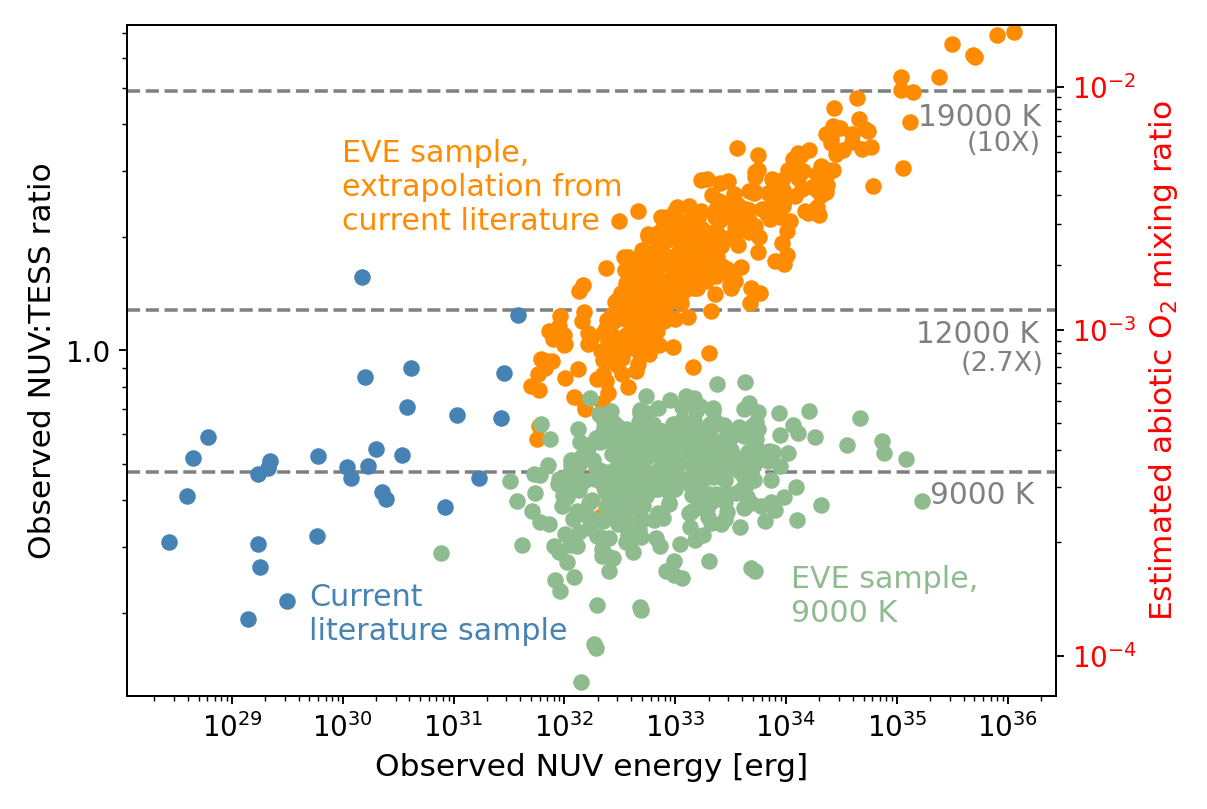}
  \end{minipage}\hfill
  \begin{minipage}[c]{0.41\textwidth}
        \caption{Simultaneous EVE observations of large flares will discriminate between constant UV—optical ratios and UV—optical ratios that increase with energy. The current literature sample of simultaneous detections from \citet{Paudel:2024} is composed of smaller flares, while EVE is sensitive to some of the largest flares emitted by young stars. The EVE samples for constant and increasing $T_\mathrm{eff}$--energy scenarios are drawn from the FFD and flare yield predictions of \S \ref{applic_eve}. The right y-axis shows the surface mixing ratio of O$_2$ in an Earth-like atmosphere resulting from photochemistry, where we assume for purposes of illustration that the flare $T_\mathrm{eff}$ that produces the observed NUV:TESS ratio can be used to estimate a FUV:NUV ratio. At the largest energies, excess UV emission corresponds to an increase of 30—60$\times$ in abiotic O2 production for Earth-like atmospheres \citep{Harman:2015}.\newline}
  \end{minipage}
  \vspace{-0.65cm}
  \label{fig:eve_ECFs}
\end{figure}

We find that the scatter in optical-NUV scaling relations can be reduced by a factor of 2.0$\pm$0.6 when constructing the relationship for the FWHM TESS energy instead of for the total energy. We do not see any qualitative differences in the energy budgets of flares from different stars, although a larger sample of stellar masses and ages is needed to confirm this observation. Future work in a larger sample is also needed to determine whether separate scaling relations for different flare morphologies are required to further reduce the presence of scatter. 

The cumulative impact of flares on the radiation environments of young planets is sensitive to both the NUV-optical scaling relation and power law index $\alpha_\mathrm{FFD}$ of the FFD. For shallow slopes with $\alpha_\mathrm{FFD}$=1.6, the cumulative radiation environment is at least 3.6$\times$ greater for the \citet{Paudel:2024} optical-NUV scaling relation than for the 9000 K one. For steep slopes of $\alpha_\mathrm{FFD}$=2.1, the cumulative radiation environment is only 1.9$\times$ greater for the \citet{Paudel:2024} optical-NUV scaling relation than for the 9000 K one. The results of \citet{Berger:2024} allow us to assume a ratio $\mathcal{R}_\mathrm{FUV/NUV}$ of 0.46$\pm$0.23 for the flares detected by EVE. In this way, EVE will simultaneously obtain the emission at photochemically-active wavelengths and optical wavelengths, enabling the exploration of more accurate optical-UV scaling relations. We note that multiple instruments used for transit spectroscopy on JWST fully cover the TESS band, allowing the TESS band energy of flares that occur during the transit to be measured \citep{Howard:2023}. The photochemical radiation environment can therefore be estimated for these flares and used for more accurate atmospheric retrievals. 

The insights gained from more accurate optical-UV scaling relations will also improve our understanding of the photochemical environments of planets to be observed with both direct imaging and transit spectroscopy by the Habitable Worlds Observatory \citep{Harada:2024}. The increase in effective flare temperature with energy observed in small flares ($E_\mathrm{NUV}<$5$\times$10$^{31}$ erg) by \citet{Paudel:2024} may extend to larger flares. Extrapolation of the $g^{'}$–TESS band optical energy ratios of 42 large flares (10$^{32}$ $< E_{g’}<$ 10$^{35}$ erg) observed simultaneously with Evryscope and TESS \citep{Howard:2020} into the NUV produces values largely in agreement with those reported for the small flares in \citet{Paudel:2024}, leading to much higher ECFs than indicated by the \citet{Jackman:2023} results. If the ECFs of large flares continue to increase with energy, our ability to infer a biological origin for O$_2$ in an Earth-like atmosphere is likely to be significantly impacted. For example, a 1D photochemical model of an Earth-like atmosphere finds the ratio of FUV:NUV emission corresponding to an ECF of 15 in the NUV produces 20$\times$ more abiotic O$_2$ than does the FUV:NUV ratio for an ECF of 2.7 and 60$\times$ more abiotic O$_2$ than does the FUV:NUV ratio for the canonical 9000 K flare blackbody \citep{Harman:2015}. EVE will reject or confirm the existence of high UV to optical ratios for these large events because it is primarily sensitive to flares with energies greater than 2$\times$10$^{32}$ due to the stellar distance to members of candidate moving groups (Figure \ref{fig:eve_ECFs}). EVE's yield of 450$\pm$100 simultaneous flare events with energies of $E_\mathrm{NUV}>$10$^{32}$ erg given a two year mission and 50\% margin will discriminate at the 3$\sigma$ level between power law slopes from 0--0.2 in increments of 0.1 for the UV–optical ratio versus energy relation. The yield would be even larger given 80\% science time and a three year mission, with 1100$\pm$160 simultaneous events.

The individual segments of simultaneous monitoring are stitched together into a single light curve of 0.6 d in order to determine to what degree the addition of the NUV band can improve the sensitivity to transits of active stars. For the first time, we demonstrate the independent information on the flare contamination provided by the NUV band can improve the transit detection efficiency by $\Delta\delta$=0.0052. For the case of a young K7 dwarf of $T$=9.64 and 0.64$R_\odot$, this increase in transit depth corresponds to an increase in the 50\% detection probability of 0.42$R_\oplus$. As a result, the 50\% detection threshold crosses the sub-Neptune to super-Earth boundary and improves from 2.19 to 1.76$R_\oplus$ at short orbital periods ($P_\mathrm{orb}\leq$1.7 d). Crucially, although the process of fitting and removing flares in single-band light curves may be susceptible to fitting and removing real transits, the independent information provided by the NUV band mitigates this concern. The approach we present for predicting the flare contribution to the TESS light curve is designed with an eye toward large-scale implementation for thousands of light curves due to its reliance on simple model fits and parallelized application for a relatively low computational cost. Future work is needed to implement joint modeling of the predicted flare structure into the Gaussian Process frameworks currently used to detrend starspot variability.

We have demonstrated the feasibility of our stellar flare science case for the EVE mission by both reviewing and advancing the state of the art for optical-UV relations and demonstrating that flares can be modeled and removed using simultaneous observations. Our trade study provides guidelines for a mission performing joint NUV and optical flare monitoring. For apertures of at least 20 cm, 30 s cadence, and 12 months of science time, we conservatively recommend designs with a FoV greater than 10 deg$^2$, and/or a combined dark current and read noise less than 15$e^{-}$ per exposure in order to distinguish populations of 9000 K and UV-luminous flares at the 3$\sigma$ level.

\section{Acknowledgements}\label{sec:acknowledge}
We would like to thank the anonymous referee for reviewing and improving the work. WH thanks Rishi Paudel, Andrew Mann, Sydney Vach, and Alexander Brown for helpful conversations on transits, extinction, and flares. Funding for this work was provided by NASA through the NASA Hubble Fellowship grant HST-HF2-51531 and HST-HF2-51530 awarded by the Space Telescope Science Institute, which is operated by the Association of Universities for Research in Astronomy, Inc., for NASA, under contract NAS5-26555. LDV acknowledges funding support from the Heising-Simons Astrophysics Postdoctoral Launch Program and from NASA under award number 80GSFC21M0002.  This work was carried out in part at the Jet Propulsion Laboratory, California Institute of Technology, under contract 80NM00018D0004 with NASA.

The 20 s TESS data from the two multi-wavelength programs was obtained for TESS Guest Observer programs 5070 (PI: MacGregor), 5121 (PI: Tovar Mendoza), 5123 (PI: Kiman), 5126 (PI: Jackman), G05144 (PI: Huber), 5145 (PI: Pietras), and 5152 (PI: Cloutier).

\par We acknowledge the use of public data from the \text{Swift} data archive. This paper includes data collected by the TESS mission. Funding for the TESS mission is provided by the NASA Explorer Program. This work has made use of data from the European Space Agency (ESA) mission {\it Gaia} (\url{https://www.cosmos.esa.int/gaia}), processed by the {\it Gaia} Data Processing and Analysis Consortium (DPAC, \url{https://www.cosmos.esa.int/web/gaia/dpac/consortium}). Funding for the DPAC has been provided by national institutions, in particular the institutions participating in the {\it Gaia} Multilateral Agreement. This research made use of Astropy,\footnote{http://www.astropy.org} a community-developed core Python package for Astronomy \citep{astropy:2013, astropy:2018,astropy:2022}, and the NumPy, SciPy, and Matplotlib Python modules \citep{numpyscipy,2020SciPy-NMeth,matplotlib}.

\software{\texttt{astropy} \citep{astropy:2013,astropy:2018,astropy:2022}}

\bibliography{References.bib}
\end{document}